%% file: arxiv.tex
\documentclass[11pt]{article}

\input{doc/macro}

\usepackage[top=2cm, bottom=2cm, left=2cm, right=2cm, includefoot]{geometry}
\usepackage{authblk}
\usepackage{graphicx}
\usepackage{amsthm}
\usepackage{url}

\begin{document}

\title{Linear Time Online Algorithms for Constructing Linear-size Suffix Trie}

\author[1]{Diptarama~Hendrian}
\author[2]{Takuya~Takagi}
\author[3]{Shunsuke~Inenaga}
\author[4]{Keisuke~Goto}
\author[3]{Mitsuru~Funakoshi}

\affil[1]{Graduate School of Information Sciences, Tohoku University, Sendai, Japan}
\affil[2]{Fujitsu Laboratories Ltd., Kawasaki, Japan}
\affil[3]{Department of Informatics, Kyushu University, Fukuoka, Japan}
\affil[4]{Independent Researcher}

\date{}

\maketitle            

\begin{abstract}
\input{doc/abstract}
\end{abstract}

\input{doc/intro}
\input{doc/preliminaries}

\input{doc/slinktree}
\input{doc/rtol}

\input{doc/rtolcomp}
\input{doc/ltor}
\input{doc/ltorcomp}
\input{doc/conclusion}

\bibliographystyle{plainurl}
\bibliography{ref}

\end{document}

%% file: doc/macro.tex
\usepackage{a4}
\usepackage{color}
\usepackage{amssymb}
\usepackage[ruled,linesnumbered,algo2e,vlined]{algorithm2e}
\usepackage{amsmath}
\usepackage{amsmath,amssymb}
\usepackage{microtype}
\usepackage{hyperref}
\usepackage{cleveref}
\usepackage{wrapfig}
\usepackage{amsthm}
\usepackage{wrapfig}
\usepackage{lineno}

\usepackage[final,mode=multiuser]{fixme}

\fxuseenvlayout{colorsig}
\fxusetargetlayout{color}

\FXRegisterAuthor{dh}{adh}{DH}
\FXRegisterAuthor{tt}{att}{TT}
\FXRegisterAuthor{si}{asi}{SI}
\FXRegisterAuthor{mf}{amf}{MF}
\fxsetface{env}{}

\newtheorem{theorem}{Theorem}
\newtheorem{lemma}{Lemma}

\newtheorem{definition}{Definition}
\newtheorem{proposition}{Proposition}
\newtheorem{observation}{Observation}

\newcommand{\etal}{et al.}

\newcommand{\STree}{\mathsf{STree}}
\newcommand{\STrie}{\mathsf{STrie}}
\newcommand{\LST}{\mathsf{LST}}
\newcommand{\PLST}{\mathsf{preLST}}
\newcommand{\SLT}{\mathbf{ELT}}

\newcommand{\mcT}{\mathcal{T}}
\newcommand{\hmcT}{\widehat{\mcT}}

\newcommand{\NMA}{\mathbf{NMA}}
\newcommand{\Mark}{\mathbf{mark}}
\newcommand{\AddLeaf}{\mathbf{addLeaf}}
\newcommand{\Demote}{\mathbf{demoteMark}}

\newcommand{\activeNode}{\mathit{activePoint}}

\newcommand{\nextNode}{\mathit{insertPoint}}

\newcommand{\Root}{\mathit{root}}
\newcommand{\newLeaf}{\mathit{newLeaf}}
\newcommand{\PrevLeaf}{\mathit{prevLeaf}}
\newcommand{\PrevInsPoint}{\mathit{prevInsPoint}}
\newcommand{\PrevLabel}{\mathit{prevLabel}}
\newcommand{\Flag}{\mathit{mismatch}}
\newcommand{\LastNode}{\mathit{newNode}}

\newcommand{\Link}{\mathbf{link}}
\newcommand{\Rlink}{\mathsf{rlink}}
\newcommand{\FLink}{\mathsf{fastLink}}
\newcommand{\Slink}{\mathsf{slink}}

\newcommand{\Tree}{\mathcal{T}}
\newcommand{\PTree}{\mathcal{P}}

\newcommand{\STriedepth}{\mathsf{STrieDepth}}
\newcommand{\Depth}{\mathsf{depth}}

\newcommand{\Child}{\mathsf{child}}
\newcommand{\TChild}{\mathsf{t1child}}
\newcommand{\Parent}{\mathsf{parent}}
\newcommand{\TParent}{\mathsf{t1parent}}

\newcommand{\CreateTypeTwo}{\mathsf{createType2}}

\newcommand{\FastMatching}{\mathsf{fastMatching}}
\newcommand{\FastDecompact}{\mathsf{fastDecompact}}
\newcommand{\ReadEdge}{\mathsf{readEdge}}
\newcommand{\Split}{\mathsf{split}}

\newcommand{\NULL}{\mathsf{NULL}}
\newcommand{\Plus}{\mathsf{+}}
\newcommand{\Type}{\mathsf{type}}
\newcommand{\Label}{\mathsf{label}}

\newcommand{\True}{\mathsf{true}}
\newcommand{\False}{\mathsf{false}}

\newcommand{\ot}{:=}

\newcommand{\Ret}{\mathbf{return}}

\newcommand{\Leaf}{\mathit{leaf}}

\SetKwProg{Fn}{Function}{}{end}

%% file: doc/abstract.tex
The suffix trees are fundamental data structures for various kinds of string processing.
The suffix tree of a text string $T$ of length $n$ has $O(n)$ nodes and edges,
and the string label of each edge is encoded by a pair of positions in $T$.
Thus, even after the tree is built, the input string $T$ needs to be kept stored
and random access to $T$ is still needed. 
The \emph{linear-size suffix tries} (\emph{LSTs}), proposed by Crochemore et al.
[Linear-size suffix tries, TCS 638:171-178, 2016],
are a ``stand-alone'' alternative to the suffix trees.
Namely, the LST of an input text string $T$ of length $n$ occupies $O(n)$ total space,
and supports pattern matching and other tasks with the same efficiency as the suffix tree
without the need to store the input text string $T$.
Crochemore et al. proposed an \emph{offline} algorithm which transforms
the suffix tree of $T$ into the LST of $T$ in $O(n \log \sigma)$ time and $O(n)$ space,
where $\sigma$ is the alphabet size.
In this paper, we present two types of \emph{online} algorithms
which ``directly'' construct the LST, from right to left, and from left to right,
without constructing the suffix tree as an intermediate structure.
Both algorithms construct the LST incrementally when a new symbol is read,
and do not access the previously read symbols.
Both of the right-to-left construction algorithm and the left-to-right construction algorithm work in $O(n \log \sigma)$ time and $O(n)$ space.
The main feature of our algorithms is that the input text string does not need to be stored.

%% file: doc/intro.tex
\section{Introduction}
Suffix tries are conceptually important text string data structures
that are the basis of more efficient data structures.
While the suffix trie of a text string $T$ supports fast queries and operations
such as pattern matching, the size of the suffix trie can be $\Theta(n^2)$ in the worst case,
where $n$ is the length of $T$.
By suitably modifying suffix tries, 
we can obtain linear $O(n)$-size string data structures 
such as suffix trees~\cite{Weiner1973},
suffix arrays~\cite{Manber1993},
directed acyclic word graphs (DAWGs)~\cite{Blumer1985},
compact DAWGs (CDAWGs)~\cite{Blumer1987},
position heaps~\cite{Ehrenfeucht2011}, and so on.
In the case of the integer alphabet of size polynomial in $n$,
all these data structures can be constructed in $O(n)$ time and space
in an \emph{offline} manner~\cite{Crochemore1997const,Crochemore1997,Farach-ColtonFM00,FujishigeTIBT16,INIBT16,KarkkainenSB06,NarisawaHIBT17}.
In the case of a general ordered alphabet of size $\sigma$,
there are \emph{left-to-right} \emph{online} construction algorithms 
for suffix trees~\cite{Ukkonen1995},
DAWGs~\cite{Blumer1985}, CDAWGs~\cite{InenagaHSTAMP05}, and position heaps~\cite{Kucherov2013}.
Also, there are \emph{right-to-left} \emph{online} construction algorithms 
for suffix trees~\cite{Weiner1973} and position heaps~\cite{Ehrenfeucht2011}.
All these online construction algorithms run in $O(n \log \sigma)$ time with $O(n)$ space.

Suffix trees are one of the most extensively studied string data structures,
due to their versatility.
The main drawback is, however, that
each edge label of suffix trees needs to be encoded as a pair of text positions,
and thus the input string needs to be kept stored and be accessed even after
the tree has been constructed.
Crochemore et al.~\cite{Crochemore2016} proposed a new suffix-trie
based data structure called \emph{linear-size suffix tries} (\emph{LSTs}).
The LST of $T$ consists of the nodes of the suffix tree of $T$,
plus a linear number of auxiliary nodes and suffix links.
Each edge label of LSTs is a single character,
and hence the input text string can be discarded after the LST has been built.
The total size of LSTs is linear in the input text string length,
yet LSTs support fundamental string processing queries such as pattern matching
within the same efficiency as their suffix tree counterpart~\cite{Crochemore2016}.

Crochemore et al.~\cite{Crochemore2016} showed
an algorithm which transforms the \emph{given} suffix tree of text string $T$
into the LST of $T$ in $O(n \log \sigma)$ time and $O(n)$ space.
This algorithm is \emph{offline}, since it requires the suffix tree to be completely built first.
No efficient algorithms which construct LSTs \emph{directly} (i.e. without suffix trees)
and in an \emph{online} manner were known.

This paper proposes two online algorithms that construct the LST directly from the given text string.
The first algorithm is based on Weiner's suffix tree construction~\cite{Weiner1973},
and constructs the LST of $T$ by scanning $T$ \emph{from right to left}.
On the other hand, the second algorithm is based on Ukkonen's suffix tree construction~\cite{Ukkonen1995},
and constructs the LST of $T$ by scanning $T$ from \emph{left to right}.
Both algorithms construct the LST incrementally when a new symbol is read,
and do not access the previously read symbols.
This also means that our construction algorithms do not need to store the input text string,
and the currently processed symbol in the text can be immediately discarded
as soon as the symbol at the next position is read.
Moreover, our algorithms also construct data structures for fast links directly,
which is necessary to perform pattern matching efficiently.
Both of the right-to-left construction algorithm and the left-to-right construction algorithm work in $O(n \log \sigma)$ time and $O(n)$ space.

In the preliminary version~\cite{Hendrian2019} of this work,
the fast links of the LST in its left-to-right construction were not explicitly created, but instead we used Alstrup et al.'s~\cite{Alstrup1998} fully-dynamic nearest marked ancestor (NMA) data structure for simulating the fast links.
Since their data structure requires $O(\log n / \log \log n)$ time for updates, the previous left-to-right LST construction algorithm takes $O(n \log n / \log \log n)$ total time for maintaining a representation of the fast links~\cite{Hendrian2019}.
In our new algorithm for left-to-right online LST construction, we use a new data structure for (limited) dynamic NMA queries on a tree of (reversed) suffix links, which suffice for us to maintain our fast links, in $O(n)$ total time and space.
A key observation of our method is that, in our representation of fast links, whenever a marked node gets unmarked, then all of its children get marked.
We show how to build such a (limited) dynamic NMA data structure by maintaining a linear-size copy of the suffix link tree.

The rest of the paper is organized as follows:
Section~\ref{sec:preliminaries} is devoted to basic definitions and notations.
In Section~\ref{sec:ELT}, we present our new data structure for (limited) dynamic NMA queries, which will then be used for our left-to-right online construction of the LST.
In Section~\ref{sec:right-to-left}, we propose our Weiner-type right-to-left online construction algorithm for the LST.
Section~\ref{sec:left-to-right} presents our Ukkonen-type left-to-right online construction algorithm for the LST.
Finally, Section~\ref{sec:conclusions} concludes and lists some open problems.

%% file: doc/preliminaries.tex
\section{Preliminaries} \label{sec:preliminaries}

Let $\Sigma$ denote an \emph{alphabet} of size $\sigma$.
An element of $\Sigma^*$ is called a \emph{string}.
For a string $T \in \Sigma^*$, the length of $T$ is denoted by $|T|$.
The \emph{empty string}, denoted by $\varepsilon$, is the string of length $0$.
For a string $T$ of length $n$, $T[i]$ denotes the $i$-th symbol of $T$
and $T[i:j] = T[i]T[i+1] \cdots T[j]$ denotes the \emph{substring} of $T$ that begins at position $i$ and ends at position $j$ for $1 \leq i \leq j \leq n$.
Moreover, let $T[i:j] = \varepsilon$ if $i > j$.
For convenience, we abbreviate $T[1:i]$ to $T[:i]$ and $T[i:n]$ to $T[i:]$,
which are called \emph{prefix} and \emph{suffix} of $T$, respectively.

\subsection{Linear-size suffix tries}
The \emph{suffix trie} $\STrie(T)$ of a string $T$ is a trie
that represents all suffixes of $T$.
The \emph{suffix link} of each node $U$ in $\STrie(T)$
is an auxiliary link that points to $V = U[2:|U|]$. 
The \emph{suffix tree}~\cite{Weiner1973} $\STree(T)$ of $T$ is a path-compressed trie
that represents all suffixes of $T$.
We consider the version of suffix trees where
the suffixes that occur twice or more in $T$
can be explicitly represented by nodes.
The \emph{linear-size suffix trie} $\LST(T)$ of a string $T$,
proposed by Crochemore et al.~\cite{Crochemore2016},
is another kind of tree that represents all suffixes of $T$,
where each edge is labeled by a single symbol.
The nodes of $\LST(T)$ are a subset of the nodes of $\STrie(T)$,
consisting of the two following types of nodes:
\begin{enumerate}
	\item Type-1: The nodes of $\STrie(T)$ that are also nodes of $\STree(T)$.
	\item Type-2: The nodes of $\STrie(T)$
              that are not type-1 nodes and whose suffix links point to type-1 nodes.
\end{enumerate}
A non-suffix type-1 node has two or more children and a type-2 node has only one child.
When $T$ ends with a unique terminate symbol $\$$ that
does not occur elsewhere in $T$,
then all type-1 non-leaf nodes in $\LST(T)$ have two or more children.
The nodes of $\STrie(T)$ that are neither type-1 nor type-2 nodes of $\LST(T)$
are called \emph{implicit nodes} in $\LST(T)$.

\begin{figure}[t]
	\centering
	\begin{minipage}[t]{0.32\hsize}
		\centering
		\includegraphics[scale=1.1]{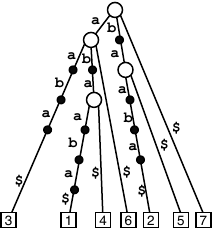}\\
		\ \ \ \small{Suffix trie}
	\end{minipage}
	\begin{minipage}[t]{0.32\hsize}
		\centering
		\includegraphics[scale=1.1]{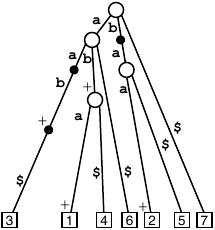}\\
		\ \ \ \small{Linear-size suffix trie}
	\end{minipage}
	\begin{minipage}[t]{0.32\hsize}
		\centering
		\includegraphics[scale=1.1]{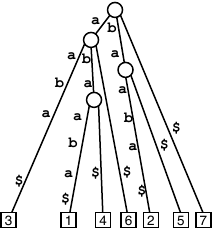}\\
		\ \ \ \small{Suffix tree}
	\end{minipage}
	\caption{
		The suffix trie, linear-size suffix trie, and suffix tree of $T = \mathtt{abaaba\texttt{\$}}$.
	}
	\label{fig:suffix_trie_tree}
\end{figure}

We identify each node in $\LST(T)$ by the substring of $T$ that is the path label from $\Root$ to the node in $\STrie(T)$.
The \emph{string depth} of a node $U$ is the length of the string that is represented by $U$.
Let $U$ and $V$ be nodes of $\LST(T)$ such that $V$ is a child of $U$.
The edge label of $(U,V) = c$ is the same as the label of the first edge on the path from $U$ to $V$ in $\STrie(T)$.
If $V$ is not a child of $U$ in $\STrie(T)$, i.e. the length of the path label from $U$ to $V$ is more than one,
we put the $\Plus$ sign on $V$ and we call $V$ a $\Plus$-node.
\Cref{fig:suffix_trie_tree} shows an example of a suffix trie, linear-size suffix trie, and suffix tree.

For convenience, we assume that there is an auxiliary node $\bot$
as the parent of the root of $\LST(T)$,
and that the edge from $\bot$ to the root is labeled by any symbol.
This assures that for each symbol appearing in $T$
the root has a non $\Plus$ child.
This will be important for the construction of LSTs 
and pattern matching with LSTs (c.f. Lemma~\ref{lem:readlabel}).

In the description of our algorithms, we will use the following notations.
For any node $U$, $\Parent(U)$ denotes the parent node of $U$.
For any edge $(U, V)$,
$\Label(U,V)$ denotes the label of the edge connecting $U$ and $V$.
For a node $U$ and symbol $c$,
$\Child(U,c)$ denotes the child of $U$ whose incoming edge label is $c$,
if it exists.
We denote $\Plus(U)=\True$ if $U$ is a $\Plus$-node, and $\Plus(U)=\False$ otherwise.
The suffix link of a node $U$ is defined as $\Slink(U)=V$, where $V = U[2:|U|]$.
The reversed suffix link of a node $V$ with a symbol $c \in \Sigma$
is defined as $\Rlink(V,c)=U$, if there is a node $V$ such that $cV = U$.
It is undefined otherwise.
For any type-1 node $U$,
$\TParent(U)$ denotes the nearest type-1 ancestor of $U$,
and $\TChild(U,c)$ denotes the nearest type-1 descendant of $U$ on $c$ edge.
For any type-2 node $U$,
$\Child(U)$ is the child of $U$,
and $\Label(U)$ is the label of the edge connecting $U$ and its child.

\subsection{Pattern matching using linear-size suffix tries}
\label{subsec:matching}
In order to efficiently perform pattern matching on LSTs,
Crochemore \etal{}~\cite{Crochemore2016} introduced \emph{fast links} that are
a chain of \emph{suffix links of edges}.

\begin{definition}
	For any edge $(U,V)$, let $\FLink(U,V)=(\Slink^h(U),\Slink^h(V))$
	such that $\Slink^{h}(U) \ne \Parent(\Slink^{h}(V))$ and $\Slink^{h-1}(U) = \Parent(\Slink^{h-1}(V))$, where $\Slink^0(U) = U$ and  $\Slink^i(U) = \Slink(\Slink^{i-1}(U))$.
\end{definition}
Here, $h$ is the minimum number of suffix links that we need to  
traverse so that $\Slink^{h}(U) \ne \Parent(\Slink^{h}(V))$.
Namely, after taking $h$ suffix links from edge $(U,V)$,
there is at least one type-2 node in the path from $\Slink^h(U)$ to $\Slink^h(V)$.
Since type-2 nodes are not branching,
we can use the labels of the type-2 nodes in this path to retrieve
the label of the edge $(U, V)$ (see \Cref{lem:readlabel} below).
Provided that $\LST(T)$ has been constructed,
the fast link $\FLink(U,V)$ for every edge $(U,V)$ can be computed
in a total of $O(n)$ time and space~\cite{Crochemore2016}.

\begin{lemma}[\cite{Crochemore2016}]\label{lem:readlabel}
  The underlying label of a given edge $(U,V)$ of length $\ell$ can be
  retrieved in $O(\ell\log \sigma)$ time by using fast links.
\end{lemma}

\subsubsection{Notes on \textsc{FastSearch} pattern matching algorithm}

Crochemore et al.~\cite{Crochemore2016} claimed that one can perform pattern matching for a given pattern $P$ in $O(|P|\log \sigma)$ time with the LST by using the algorithm \textsc{FastSearch}.
This algorithm, however, is not described in detail in~\cite{Crochemore2016}, even if the ideas present in~\cite{Crochemore2016} seem correct.
Moreover, the proof provided in~\cite{Crochemore2016} for time efficiency of their pattern matching algorithm seems incomplete,
because the proof by Crochemore et al.~\cite{Crochemore2016} does not explicitly describe the case where pattern matching terminates on an edge that has a long underlying label.
That is, what is missing in the proof in~\cite{Crochemore2016} is that
the number of applications of fast links (and thus the number of recursive calls) is bounded by the pattern length $m$ for such a case.
Due to this, for searching a suffix (i.e. the last fragment) of a given pattern $P$ of length $s$ for an edge $(U,V)$ of length $\ell$,
what one can guarantee with the original proof in~\cite{Crochemore2016} is mere $O(\ell)$ pattern matching running time, even when $s < \ell$.

The next lemma describes an instance where the length $\ell$ of the last edge can be arbitrarily long and cannot be bounded by the pattern length $m$:

\begin{figure}[t]
        \centering
	\includegraphics[scale=0.35]{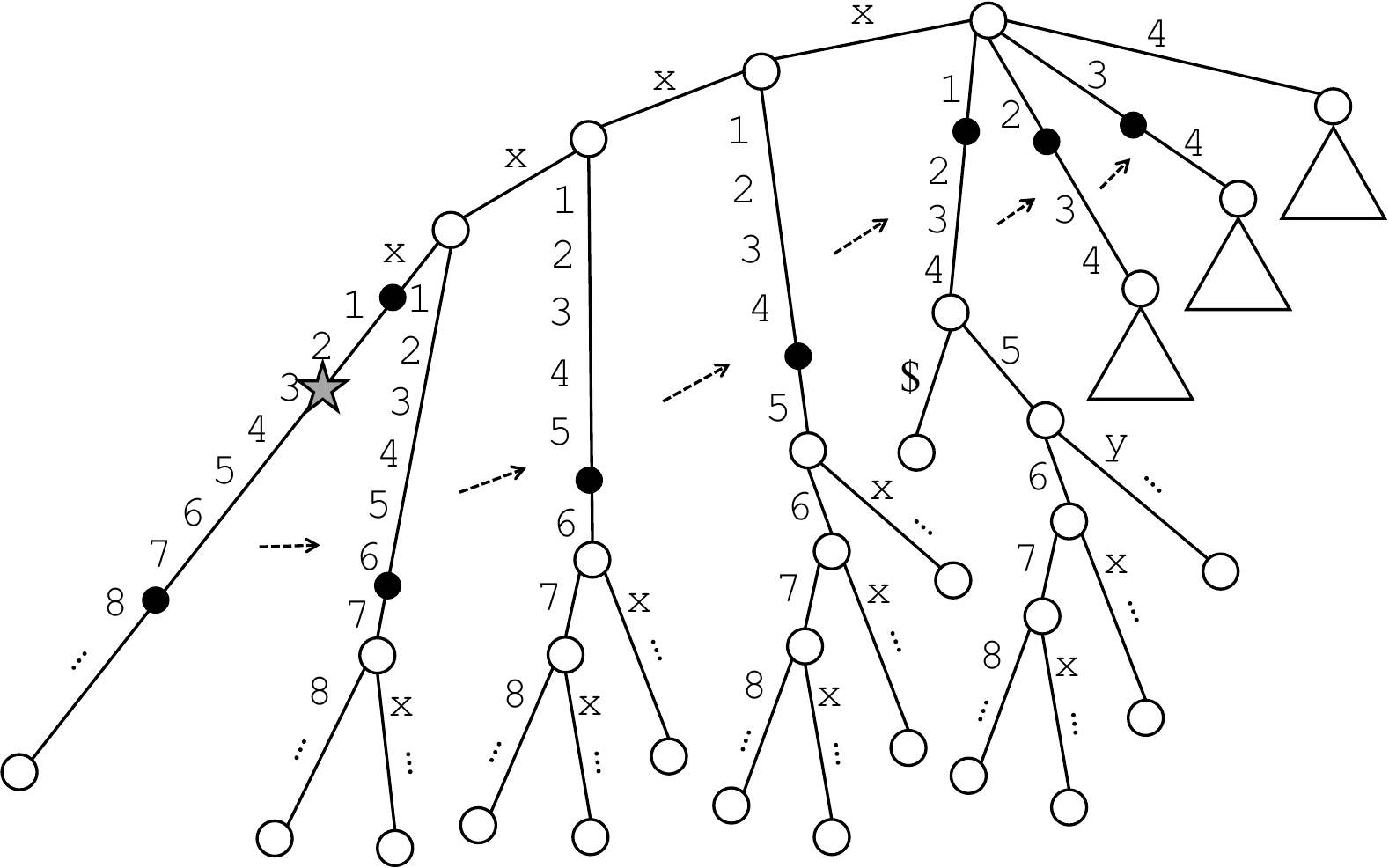}
	\caption{Illustration that shows a relevant part of $\LST(T)$ for the string $T = \mathtt{xxxx} \mathtt{12345678}\mathtt{xxx} \mathtt{1234567} \mathtt{xx} \mathtt{123456} \mathtt{x} \mathtt{12345} \mathtt{y1234}\$$ from \Cref{lem:lowerbound}, where we have set the parameters $m = 4$ and $k = 8$ in this concrete example. For an illustrative purpose, some string edge labels are explicitly shown. The dashed arrows represent fast links. For instance, when pattern $P = \mathtt{xxxx12}$ of length $m+2$ is given (its locus is designated by the star symbol), then it must be guaranteed that the number of applications of fast links for matching $P$ is bounded by $O(m)$ and is independent of the length $k-1 = \Theta(n/m)$ of the last edge labeled $\mathtt{1234567}$.}
	\label{fig:lowerbound}
\end{figure}

\begin{proposition} \label{lem:lowerbound}
  There exists a family of strings $T$ of length $n$ for which $\LST(T)$ contains
  an edge $(U, V)$ such that node $U$ is of string depth $\Theta(m)$,
  and the length of the underlying label of the edge $(U, V)$ is $\Theta(n/m)$.
\end{proposition}
\begin{proof}
  Consider the following string $T$ over the alphabet $\Sigma = \{\mathtt{x,y}, \$\} \cup \{\mathtt{1}, \ldots, \mathtt{k}\}$ of size $k+3$,
  \[
  T = (\mathtt{x}^m \mathtt{1} \cdots \mathtt{k}) (\mathtt{x}^{m-1} \mathtt{1} \cdots (\mathtt{k-1})) \cdots (\mathtt{x}^{1} \mathtt{1} \cdots (\mathtt{k-m+1})) (\mathtt{y} \mathtt{1} \cdots (\mathtt{k-m}))\$,
  \]
  where $k-m = \Theta(k)$ (and thus $m < k$). Note that $|T| = n = \Theta(mk)$.
  Then, the substring $\mathtt{x}^{m}$ is represented by a type-2 node,
  and let denote it by $U$.
  Let $(U, V)$ be the out-edge of $U$ whose underlying edge label
  is $\mathtt{1} \cdots (\mathtt{k-1})$ of length $k-1$
  (see also Figure~\ref{fig:lowerbound}).
  Then, it is clear that the string depth of $U$ is exactly $m$
  and the length of the underlying label of the edge $(U, V)$ is
  $k-1 = \Theta(n/m)$.
\end{proof}

Due to \Cref{lem:lowerbound}, it needs to be clarified that pattern matching queries can be supported only with $O(m)$ applications of fast links, irrespectively the length $\ell$ of the last edge.

\subsubsection{Full proof for linear-time pattern matching with LST}

In the rest of this subsection, we describe an algorithm which efficiently performs the longest prefix match (not just the simple search) for a given pattern on the LST with fast links. The proposed algorithm is described in the following lemma.

\begin{lemma}\label{lem:patmatch}
	Given $\LST(T)$ and a pattern $P$, 
	we can find the longest prefix $P'$ of $P$ that occurs in $T$ in $O(|P'|\log\sigma)$ time.
\end{lemma}

\begin{figure}[b]
        \centering
	\includegraphics[scale=0.55]{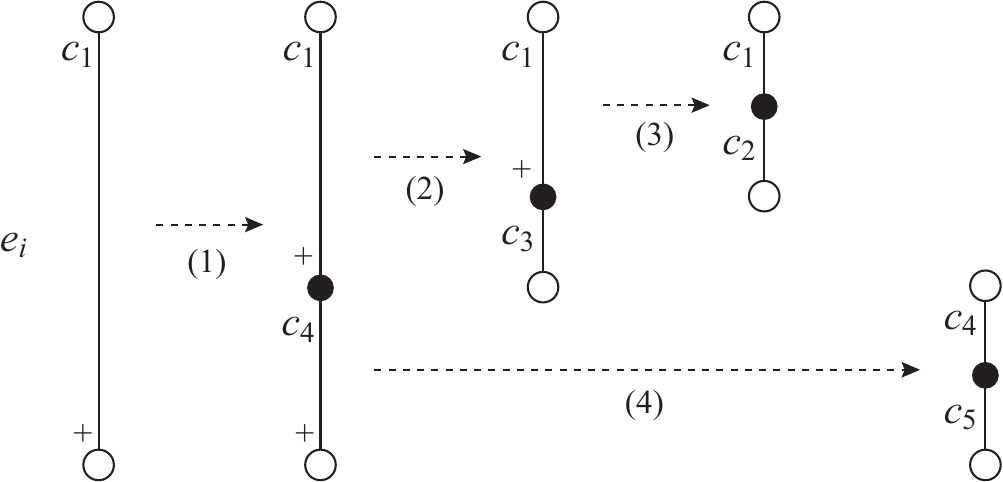}
	\caption{Illustration for our pattern matching algorithm with LST. The dashed arrows represent fast links. The number in parentheses show the orders of applications of fast links when traversing $P_i = c_1 c_2 c_3 c_4 c_5$ on the edge $e_i$.}
	\label{fig:pattern_matching_1}
\end{figure}

\begin{figure}[t]
        \centering
	\includegraphics[scale=0.55]{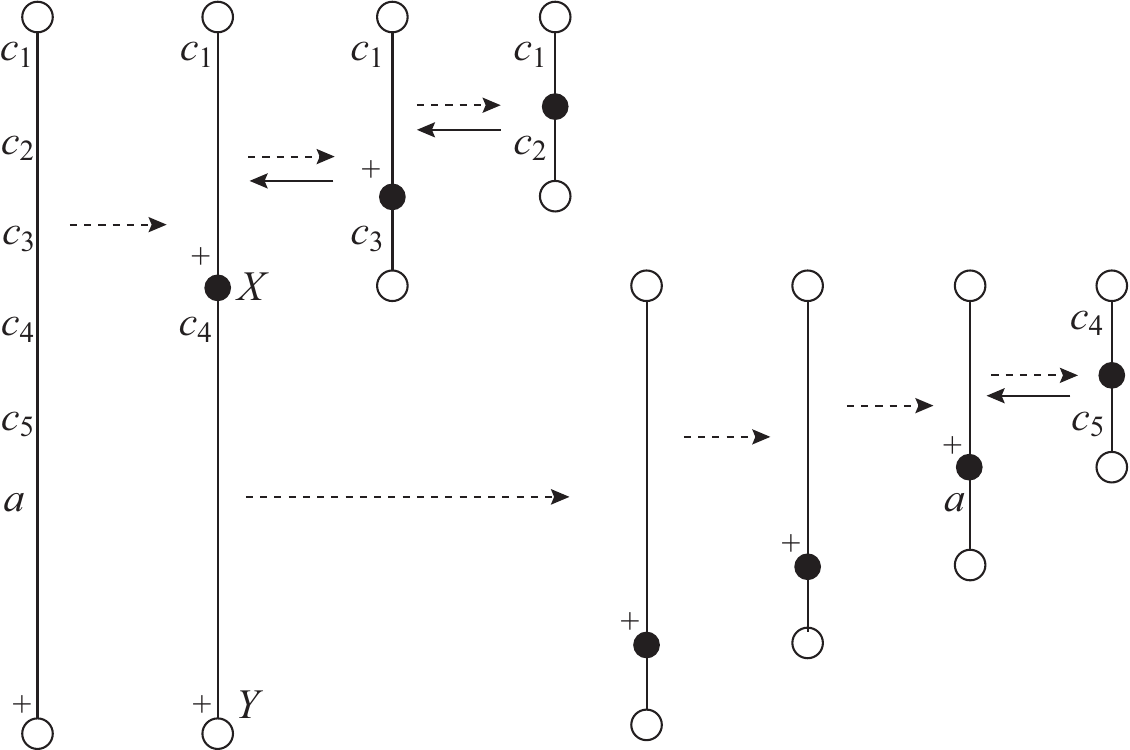}
	\caption{Illustration for a retrieval of the last fragment $P_m$ of longest prefix $P' = P_1 \cdots P_m$, where $P_m = c_1 c_2 c_3 c_4 c_5$ and $a$ is the first mismatching symbol in this figure. The dashed arrows represent fast links, and the solid arrows show the fast links that are traced back. The non-traced back fast links are in the last part of our successive applications of fast links. If it started from edge $(X, Y)$, then the number of such non-traced back fast links is bounded by the string depth of node $X$, which is at most $|P'|$.}
	\label{fig:pattern_matching_2}
\end{figure}

\begin{proof}
  Below we consider the case where the pattern matching with $P$
  terminates with a mismatch. The case where $P$ is a substring of $T$ is analogous.

  Let $P_{1}P_{2} \cdots P_{m} = P'$ be the factorization of $P'$ such that $P_{1} \cdots P_{i}$ is represented by a node in $\LST(T)$ for $1 \le i < m$,  $P_{1} \cdots P_{i} = \Parent(P_{1} \cdots P_{i+1})$ for $1 \le i < m-1$, and $P_{1} \cdots P_{m-1}$ is the longest prefix of $P'$ that is represented by a node in $\LST(T)$. If $P_{1} \cdots P_{m-1} = P'$, then $P_m = \varepsilon$. In what follows, we consider a general case where $P_m \neq \varepsilon$.

  Suppose that we have successfully traversed up to $P_1 \cdots P_{i-1}$,
  and let $U$ be the node representing $P_1 \cdots P_{i-1}$.
  Now we are going to traverse $P_i = c_1 \cdots c_{|P_i|}$.
  If $U$ has no outgoing edge labeled $c_1 = P_i[1] = P[|P_1 \cdots P_{i-1}|+1]$,
  then the traversal terminates on $U$.
  Suppose $U$ has an outgoing edge labeled $c_1$ and let $V$ be the child
  of $U$ with the $c_1$-edge.
  We denote this edge by $e_i = (U, V)$.
  There are two cases:
  \begin{enumerate}
  \item If $V$ is not a $\Plus$-node, then $P_i = c_1$.
  We then set $U \leftarrow V$ and move on to the next factor $P_{i+1}$.
  \item Otherwise (if $V$ is a $\Plus$-node),
  then we apply $\FLink$ from edge $(U, V)$ recursively,
  until reaching the first edge $(U', V')$ such that $V'$ is not a $\Plus$-node
  (see also Figure~\ref{fig:pattern_matching_1} for illustration).
  Then we move onto $V'$. Note that by the definition of $\FLink$,
  $V'$ is always a type-2 node.
  We then continue the same procedure by setting $U \leftarrow V'$
  with the next pattern symbol $c_2$.
  This will be continued until we arrive at the first edge $(U, V)$
  such that $V$ is a type-1 node.
  Then, we trace back the chain of $\FLink$'s from $(U, V)$
  until getting back to the type-2 node $V''$
  whose outgoing edge has the next symbol to retrieve.
  We set $U \leftarrow V''$ and continue with the next symbol.
  This will be continued until we traverse all symbols $c_j$ in $P_i$
  in increasing order of $j = 1, \ldots, |P_i|$ along the edge $e_i$,
  or find the first mismatching symbol right after $P_m$.
  \end{enumerate}

  The correctness of the above algorithm follows from the fact that
  every symbol in the label of the edge $e_i$ is retrieved from a type-2 node
  that is not branching, except for the first one retrieved from
  the type-1 node that is the origin of $e_i$.
  Since any type-2 node is not branching,
  we can traverse the edge $e_i$ with $P_i$
  iff the underlying label of $e_i$ is equal to $P_i$ for $1 \leq i \leq m-1$.
  The case of the last edge $e_{m}$ where the first mismatching symbol is found is analogous.
    
  To analyze the time complexity, we consider the number of applications of $\FLink$.
  For each $1 \leq i \leq m-1$, the number of applications of $\FLink$ is
  bounded by the length of the underlying label of edge $e_i$, which is $|P_i|$.
  This is because each time we trace back a $\FLink$, the corresponding new symbol is retrieved.
  Hence we can traverse $P_1 \cdots P_{m-1}$ in $O(|P_1 \cdots P_{m-1}| \log \sigma)$ time.
  As for the last fragment $P_m$,
  the number of $\FLink$'s which are traced back is at most $|P_{m}|$.
  Thus what remains is the number of $\FLink$'s which are not traced back
  and hence do not correspond to the symbols in $P_m$.
  Note that such $\FLink$'s exist in the last part of our applications of $\FLink$ (see also Figure~\ref{fig:pattern_matching_2}).
  Let $(X, Y)$ be the edge from which the applications of $\FLink$,
  which are not traced back, started.
  Our key observation is that the string depth of the node $X$ is at most $|P'|$.
  The number of applications of $\FLink$ from the edge $(X, Y)$ is
  bounded by the number of suffix links from the node $X$ to the root,
  which is $|P'| = |P_1 \cdots P_m|$.
  Thus, the total number of applications of $\FLink$ for $P_m$ is $O(|P'|)$.
  Overall, it takes $O(|P'| \log \sigma)$ time
  to traverse $P' = P_1 \cdots P_m$. This completes the proof.
\end{proof}
\Cref{alg:patmatch} shows a pseudo-code of our pattern matching algorithm with the LST in \Cref{lem:patmatch}.
In order to be self-contained we have included in \Cref{alg:patmatch} a rewriting of function \textsc{FastDecompact}~\cite{Crochemore2016} that is compatible with all the pseudo-codes in our paper.

\begin{algorithm2e}[t]
	\caption{Fast pattern matching algorithm with LSTs}
	\label{alg:patmatch}
	\SetVlineSkip{0.5mm}
	let $P$ be a pattern and $i$ be a global index.
	
	\Fn{$\FastMatching(P)$}{
		$U \ot \Root$;
		$i \ot 1$\;
		\While{$i \le |P|$}{
			\If{$\Child(U,P[i]) \ne \NULL$}{
				$U \ot \FastDecompact(U,\Child(U,P[i]))$\;
				\lIf{$U = \NULL$}{
					$\Ret$ $\False$}
			}
			\lElse{
				$\Ret$ $\False$}
		}
		$\Ret$ $\True$\;
	}
	
	\Fn{$\FastDecompact(U,V)$}{
		\While{$U \ne V$}{
			\If{$\Child(U,P[i]) \ne \NULL$}{
				\If{$\Plus(\Child(U,P[i])) = \False$}{
					$U \ot \Child(U,P[i])$\;
					$i \ot i+1$\;
				}
				\Else{
					$(U',V') = \FLink(U,\Child(U,P[i]))$\;
					$U = \FastDecompact(U',\Child(U',P[i]))$}
				\lIf{$i > |P|$}{
					$\Ret$ $V$}
			}
			\lElse{
				$\Ret$ $\NULL$}
		}
		$\Ret$ $V$\;
	}
\end{algorithm2e}

%% file: doc/slinktree.tex
\section{Nearest Marked Ancestors on Edge Link Trees}\label{sec:ELT}

As described in \Cref{subsec:matching}, we need to use fast links of LSTs to perform pattern matching efficiently.
The fast link $\FLink(U,V)$ for every edge $(U,V)$ of $\LST(T)$ can be computed
in a total of $O(n)$ time and space \emph{offline} due to Crochemore et al.~\cite{Crochemore2016}.
However, our linear-size suffix trie $\LST(T)$ built in an online manner is a growing tree,
and the offline method by Crochemore et al.~\cite{Crochemore2016} is not applicable to the case of growing trees.

\subsection{Edge Link Trees for LSTs}
In order to maintain fast links on a growing tree,
we use \emph{nearest marked ancestor queries} on the \emph{edge link tree} of $\LST(T)$.

\begin{figure}[t]
	\centering
	\begin{minipage}[t]{0.49\hsize}
		\centering
		\includegraphics[scale=1.1]{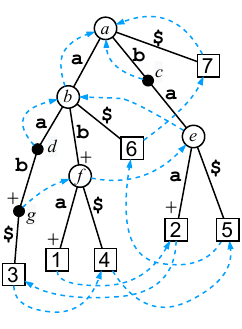}\\
		\ \ \ \small{Linear-size suffix trie}
	\end{minipage}
	\begin{minipage}[t]{0.49\hsize}
		\centering
		\includegraphics[scale=1.1]{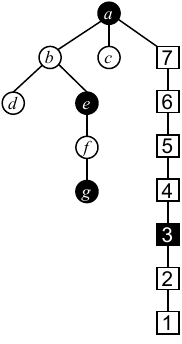}\\
		\ \ \ \small{Edge link tree}
	\end{minipage}
	\caption{
		Linear-size suffix trie of $T = \mathtt{abaaba\texttt{\$}}$ and its edge link tree.
	}
	\label{fig:lstelt}
\end{figure}

Conceptually, the edge links are suffix links of edges,
and such links can be found in the literature~\cite{LarssonFK14}
(these links are referred to as \emph{edge-oriented suffix links} therein).
On the other hand, 
since $\LST(T)$ is a tree, we can identify any edge with its destination node,
so that edge links are links from a node to a node.
This means that the edge link tree is basically the reversed suffix link tree of $\LST(T)$,
but in addition we mark some of its nodes depending on the type-2 nodes of $\LST(T)$.
Now, the edge link tree of $\LST(T)$ is formally defined as follows:
\begin{definition}[Edge link trees]
  The edge link tree $\SLT(T) = (\mathbb{V'}, \mathbb{E'})$ for the linear-size suffix trie $\LST(T) = (\mathbb{V}, \mathbb{E})$
  is a rooted tree such that $\mathbb{V'} = \mathbb{V}$
  and $\mathbb{E'} = \{(U, V) \mid U = \Slink(V) \mbox{ on } \LST(T)\}$.
  The root of $\SLT(T)$ is always marked.
  A non-root node $V \in \mathbb{V'}$ of $\SLT(T)$ is \emph{marked}
  if the parent of $V$ in $\LST(T)$ is a type-2 node,
  and $V$ is \emph{unmarked} otherwise.
\end{definition}
\Cref{fig:lstelt} shows an example of an edge link tree.

For any node $U$ in $\SLT(T)$,
let $\NMA(U)$ denote the nearest marked ancestor.
We include $U$ itself as an ancestor of $U$, so that $\NMA(U) = U$ if $U$ is marked.
We can simulate fast links using NMA queries on $\SLT(T)$ from the following observation:
\begin{observation} \label{obs:nma_fastlink}
	For an edge $(U,V)$ in $\LST(T)$, let $\FLink(U,V) = (U',V')$.
	Then, $V' = \NMA(\Slink(V))$ in $\SLT(T)$.
\end{observation}

Recall that $\TParent(V')=U'$, that is, $U'$ is the lowest type-1 ancestor of $V'$.
This means that the fast link of edge $(U,V)$ is $\FLink(U,V) = (\TParent(\NMA(\Slink(V))),\NMA(\Slink(V)))$.

For a concrete example, see Figure~\ref{fig:lstelt}.
The fast link of the edge $(f, 1)$ of the LST, which is labeled by $\mathtt{a+}$,
points to the node pair $(b, 3)$ having two type-2 nodes between them.
The NMA of node $1$ of the edge link tree is $3$, which is the destination node
of the LST edge $(g, 3)$.

Next, we show how to maintain edge link trees on a growing LST $\LST(T)$
where $T$ is updated in an online manner.
We consider the following queries and operations on the edge link tree $\SLT(T)$.
Let $U$ be a given node of $\SLT(T)$:
\begin{enumerate}
	\item $\NMA(U)$: find the nearest marked ancestor of node $U$ in $\SLT(T)$.
	\item $\Mark(U)$: mark node $U$ in $\SLT(T)$.
	\item $\AddLeaf(U)$: add a leaf as a new child of node $U$ in $\SLT(T)$.
	\item $\Demote(U)$: if node $U$ is marked, unmark $U$ and mark all children of $U$ in $\SLT(T)$.
\end{enumerate}
$\NMA(U)$ is used to implement fast links (due to Observation~\ref{obs:nma_fastlink}).
$\AddLeaf(U)$ is used when we add a new node to $\LST(T)$ and $\Mark(U)$ is used when we add a new type-2 node.
$\Demote(U)$ is used when we update a type-2 node to a type-1 node.

We can perform $\NMA(U)$, $\Mark(U)$, and $\AddLeaf(U)$ in $O(\log n)$ amortized time on a growing tree by simply using a ``relabel the smaller half'' method,
where $n$ is the length of the current string $T$.
Gabow and Tarjan~\cite{Gabow1983,Gabow1985}, also Imai and Asano~\cite{Imai1987},
proposed split-find data structures on trees by combining macro-mezzo-micro tree decomposition method and ``relabel the smaller half'' method that can be applied for NMA queries on growing trees.
By using their method, we can perform $\NMA(U)$, $\Mark(U)$, and $\AddLeaf(U)$ in $O(1)$ amortized time.
However, they do not consider $\Demote(U)$ in their method.
On the other hand, Alstrup~\etal~\cite{Alstrup1998} proposed a data structure for NMA queries on dynamic trees.
Their method supports both mark and unmark operations which can be used for $\Demote(U)$.
However, in their data structure, mark/unmark operations take $O(\log\log n)$ time and NMA queries take $O(\log n / \log \log n)$ time each.
Moreover, they showed that the time bound is tight if we use both mark and unmark operations on growing trees.

In our edge link trees, unmark operations are performed only as
a result of $\Demote(U)$, and therefore we do not need to support unmark operation on an arbitrary node in our algorithm.
This enables us to maintain an NMA data structure on our $\SLT(T)$
in linear time and space.
This will be discussed in the following subsection.

\begin{figure}[t]
	\centering
		\includegraphics[scale=0.8]{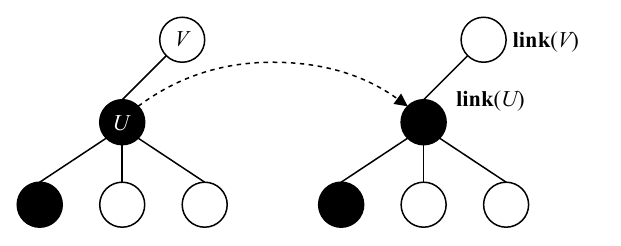}\\
		\ \ \ \small{$\mcT$ and $\hmcT$ before $\Demote(U)$}
	\\
		\includegraphics[scale=0.8]{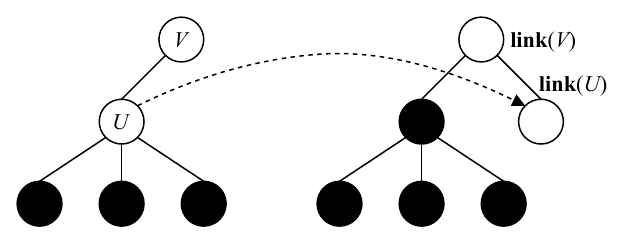}\\
		\ \ \ \small{$\mcT$ and $\hmcT$ after $\Demote(U)$}
	\caption{
		An illustration of $\Demote(U)$ on $\mathcal{T}$ (left) and its corresponding operations on $\widehat{\mathcal{T}}$ (right).
		After $\Demote(U)$, $U$ is unmarked and all children of $U$ are marked on $\mathcal{T}$.
		On the other hand, a new leaf is added to $\Link(V)$, all children of $\Link(U)$ are marked,
		and $\Link(U)$ is redirected to the new leaf on $\hmcT$.
	}
	\label{fig:eltdemote}
\end{figure}

\subsection{New data structure for NMA with demote mark on growing trees}

Here, we show that $\NMA(U)$, $\Mark(U)$, $\AddLeaf(U)$, and $\Demote(U)$ operations can be performed in $O(1)$ amortized time and linear space on a growing tree,
assuming that each operation is executed at most $n$ times, where $n$ is the final size of the tree.
To show this, we introduce an additional tree structure $\widehat{\mathcal{T}}$.
We show that we can simulate $\NMA(U)$, $\Mark(U)$, $\AddLeaf(U)$, and $\Demote(U)$ operations on $\mathcal{T}$
by using only $\NMA(\widehat{U})$, $\Mark(\widehat{U})$, and $\AddLeaf(\widehat{U})$ operations on this additional tree $\widehat{\mathcal{T}}$,
where $\widehat{U}$ is a node of $\widehat{\mathcal{T}}$.

Let $\mathcal{T}$ be an initial tree in which some nodes can be marked, and $\widehat{\mathcal{T}}$ be its copy\footnote{While the algorithm proposed in this section works for any initial tree $\mathcal{T}$, in Section~\ref{sec:left-to-right} we begin with a tree only with the root for our application of maintaining fast links in left-to-right LST construction.}.
We denote a node of $\mathcal{T}$ by $U$ and a node of $\widehat{\mathcal{T}}$ by $\widehat{U}$.
At first we link each node $U$ of $\mathcal{T}$ with its corresponding node $\widehat{U}$ in $\widehat{\mathcal{T}}$,
namely $\Link(U) = \widehat{U}$.
We consider editing $\widehat{\mathcal{T}}$ when we perform an operation on a node $U$ of $T$ as follows.
We perform $\NMA(\Link(U))$ and $\Mark(\Link(U))$,
when an operation of $\NMA(U)$ and $\Mark(U)$ performed on $U$, respectively.
Let $V$ be the new leaf when we performed $\AddLeaf(U)$, 
we perform $\AddLeaf(\Link(U))$ and $\Link(V) = \widehat{V}$,
where $\widehat{V}$ is the new leaf when we performed $\AddLeaf(\Link(U))$.
Last, we consider when we performed $\Demote(U)$.
For each child $W$ of $U$, we perform $\Mark(\Link(W))$.
Moreover, let $V$ be the parent of $U$,
we perform $\AddLeaf(\Link(V))$.
Let $\widehat{U'}$ be the new leaf, we redirect the link of $U$ to $\widehat{U'}$, namely $\Link(U)=\widehat{U'}$.

To guarantee that NMA queries on $\mathcal{T}$ can be simulated on $\hmcT$,
we need to show that $\Link(\NMA(U)) = \NMA(\Link(U))$
holds after some arbitrary operations on $\mcT$ and their corresponding operations on $\hmcT$.
First, we show the following basic relation.
\begin{lemma}\label{lem:marktt}
	For any node $U$ of $\mathcal{T}$, $\Link(U)$ is marked iff $U$ is marked.
\end{lemma}
\begin{proof}
	We prove this by induction.
	Initially $\widehat{\mathcal{T}}$ is a copy of $\mathcal{T}$, thus $\Link(U)$ is marked iff $U$ is marked.
	Assume that at a time point after some operations on $\mathcal{T}$ and $\widehat{\mathcal{T}}$ \Cref{lem:marktt} holds.
	Consider an operation on $\mathcal{T}$ and $\widehat{\mathcal{T}}$ at this point.
	After an operation $\NMA(U)$, $\Mark(U)$, or $\AddLeaf(U)$, clearly \Cref{lem:marktt} holds.
	Next, consider $\mathcal{T}$ and $\widehat{\mathcal{T}}$ after $\Demote(U)$.
	By the assumption, for any child $V$ of $U$, $\Link(V)$ is marked iff $V$ is marked.
	After $\Demote(U)$, both  $V$ and $\Link(V)$ is marked.
	Moreover, $U$ is unmarked and $\Link(U) = U'$, where $U'$ is a new leaf which is not marked.
	Therefore, \Cref{lem:marktt} holds after $\Demote(U)$.
\end{proof}
By \Cref{lem:marktt}, if a node $U$ is marked,
then $\Link(\NMA(U)) = \NMA(\Link(U))$.
In case where $U$ is not marked,
to show that $\Link(\NMA(U)) = \NMA(\Link(U))$,
we first need to show that the parent of $U$ is linked to the parent of $\Link(U)$.

\begin{figure}[t]
	\centering
	\includegraphics[scale=0.8]{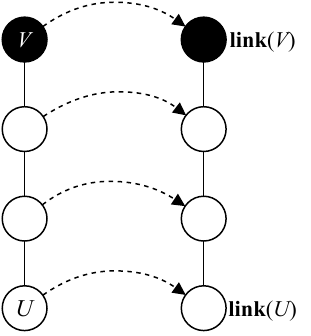}\\
	\caption{
		Any node between an unmarked node $U$ and $V = \NMA(U)$ in $\mcT$
		is linked with a node between $\Link(U)$ and $\Link(V)$ in $\hmcT$.
	}
	\label{fig:eltnma}
\end{figure}

\begin{lemma}\label{lem:parentchildtt}
	Let $U$ and $V$ be nodes of $\mathcal{T}$ such that $V$ is the parent of $U$.
	If $U$ is not marked, $\Link(V)$ is the parent of $\Link(U)$ in $\widehat{\mathcal{T}}$.
\end{lemma}
\begin{proof}
	We prove this by induction.
	Initially $\widehat{\mathcal{T}}$ is a copy of $\mathcal{T}$, 
	thus \Cref{lem:parentchildtt} holds.
	Assume that at a time point after some operations on $\mathcal{T}$ and $\widehat{\mathcal{T}}$ \Cref{lem:parentchildtt} holds.
	After an operation $\NMA(W)$, $\Mark(W)$, or $\AddLeaf(W)$ on some node $W$, clearly \Cref{lem:parentchildtt} holds,
	because $\Link(U)$ and $\Link(V)$ are not updated.
	Next, consider an operation $\Demote(W)$ on some node $W$.
	If $W \ne V$ and $W \ne U$, $\Link(U)$ and $\Link(V)$ are not updated, thus \Cref{lem:parentchildtt} holds.
	If $W = V$, then $U$ is marked, thus we do not need to consider this case.
	If $W = U$, by the definition of $\Demote(W)$,
	$\Link(W) = W'$ where $W'$ is the new leaf which is a child of $\Link(V)$
	as we can see in \Cref{fig:eltdemote}.
	Therefore, \Cref{lem:parentchildtt} holds after $\Demote(W)$.
\end{proof}
Let $U$ be an unmarked node and $V = \NMA(U)$,
for any node $W$ such that $W$ is an ancestor of $U$ and a descendant of $V$,
$\Link(W)$ is an ancestor of $\Link(U)$ and a descendant of $\Link(V)$
by \Cref{lem:parentchildtt}.
\Cref{fig:eltnma} shows this property.

Finally, by using the above property, 
we show that $\Link(\NMA(U)) = \NMA(\Link(U))$ holds after some arbitrary operations.
\begin{lemma}\label{lem:nmatt}
	For any node $U$ of $\mathcal{T}$,
	$\Link(\NMA(U)) = \NMA(\Link(U))$ holds.
\end{lemma}
\begin{proof}
	We prove this by induction.
	Initially $\widehat{\mathcal{T}}$ is a copy of $\mathcal{\mathcal{T}}$, 
	thus for any node $U$ of $T$, $\Link(\NMA(U)) = \NMA(\Link(U))$ holds.
	Assume that at a time point after some operations on $T$ and $\widehat{\mathcal{T}}$ \Cref{lem:nmatt} holds.
	Consider an operation on $\mathcal{T}$ and $\widehat{\mathcal{T}}$ at this point.
	First, we consider $\AddLeaf(U)$.
	Let $V$ be the new leaf.
	If $U$ is marked, $\Link(U)$ is also marked by \Cref{lem:marktt},
	thus $\NMA(\Link(V)) = \Link(U) = \Link(\NMA(V))$.
	
	Next, we consider $\Mark(U)$.
	For any nodes $V$ and $W$ such that $W$ is an ancestor of $V$ and is a descendant of $\NMA(V)$,
	$\Link(W)$ is an ancestor of $\Link(V)$ and is a descendant of $\Link(\NMA(V))$ by \Cref{lem:parentchildtt}.
	Thus, if $U$ is an ancestor of $V$ and is a descendant of $\NMA(V)$,
	$\NMA(V) = U$ and $\NMA(\Link(V)) = \Link(U) = \Link(\NMA(V))$.
	Otherwise, $\NMA(V)$ and $\NMA(\Link(V)) = \Link(\NMA(V))$ do not changed.
	
	Last, we consider $\Demote(U)$.
	Let $V$ and $W$ be nodes such that $\NMA(V) = W \ne U$ before $\Demote(U)$.
	Then, it is clear that $\NMA(V) = W$ and $\NMA(\Link(V)) = \Link(W) = \Link(\NMA(V))$ after $\Demote(U)$.
	Next, let $V$ be a node such that $\NMA(V) = U$ before $\Demote(U)$.
	If $U = V$, by the definition of $\Demote(U)$, $\NMA(U) = \NMA(W)$ and $\NMA(\Link(U)) = \NMA(\Link(W))$,
	where $W=\Parent(U)$.
	By the induction assumption, we have $\NMA(\Link(W)) = \Link(\NMA(W))$, thus $\NMA(\Link(U)) = \Link(\NMA(W)) = \Link(\NMA(U))$.
	Otherwise, if $U \ne V$, by \Cref{lem:parentchildtt}, there is an unmarked child $W$ of $U$ such that $W$ is an ancestor of $V$
	and $\Link(W)$ is an ancestor of $\Link(V)$.
	We note that $W = V$ can hold.
	Thus, $\NMA(V) = W$ by $\Demote(U)$ and $\NMA(\Link(V)) = \Link(W) = \Link(\NMA(V))$ by $\Demote(U)$.
\end{proof}

Last, we can show the time complexity of operations on $\mcT$ by using the time complexity of operations on $\hmcT$.
\begin{lemma}\label{lem:nmainc}
	$\NMA(U)$, $\Mark(U)$, $\AddLeaf(U)$, and $\Demote(U)$ operations can be performed in $O(1)$ amortized time on a growing tree, assuming that each operation is executed at most $n$ times, where $n$ is the final size of the tree.
\end{lemma}

\begin{proof}
	The main point of the proof is to show that the operations can be performed in $O(\log n)$ amortized time by using ``relabel the smaller half'' method.
	By using macro-mezzo-micro tree decomposition it can be reduced to $O(1)$ amortized time.
	We will prove it by using $\widehat{\mathcal{T}}$ as defined above.
	
	It is clear that $\NMA(U)$ and $\AddLeaf(U)$ can be performed in $O(1)$ time.
	$\Demote(U)$ on $\mathcal{T}$ can be reduced to 
	$\AddLeaf(\Link(\Parent(U)))$ and $\Mark(\Link(W))$
	for all children $W$ of $U$ on $\widehat{\mathcal{T}}$.
	Next, we consider the cost of $\Mark(U)$.
	Let $V = \NMA(U)$, $\widehat{U} = \Link(U)$, and $\widehat{V} = \Link(V) = \NMA(\widehat{U})$.
	By \Cref{lem:parentchildtt} and \Cref{lem:nmatt},
	the number of nodes of $\mathcal{T}$ whose NMA is $V$ is the same as the number of nodes of $\widehat{\mathcal{T}}$ whose NMA is $\widehat{V}$.
	Moreover, the number of descendants of $U$ whose NMA is $V$ is the same as the number of descendants of $\widehat{U}$ whose NMA is $\widehat{V}$.
	Thus the cost of operation $\NMA(U)$ and $\Mark(\widehat{U})$ by using ``relabel the smaller half'' method is the same.
	Therefore, the time complexity of updating $\mathcal{T}$ and $\widehat{\mathcal{T}}$ are the same which is
	$O(1)$ amortized time by using macro-mezzo-micro tree decomposition method and ``relabel the smaller half'' method.
\end{proof}

%% file: doc/rtol.tex
\section{Right-to-left online algorithm} \label{sec:right-to-left}

In this section, we present an online algorithm that constructs $\LST(T)$
by reading $T$ from right to left.
Let $\Tree_{i} = \LST(T[i:])$ for $1 \leq i \leq n$.
Our algorithm constructs $\Tree_{i}$ from $\Tree_{i+1}$ incrementally when $c = T[i]$ is read.
For simplicity, we assume that $T$ ends with a unique terminal symbol $\texttt{\$}$
such that $T[i] \ne \texttt{\$}$ for $1 \le i < n$.

\begin{figure}[t]
        \centering
	\fbox{\includegraphics[scale=0.45]{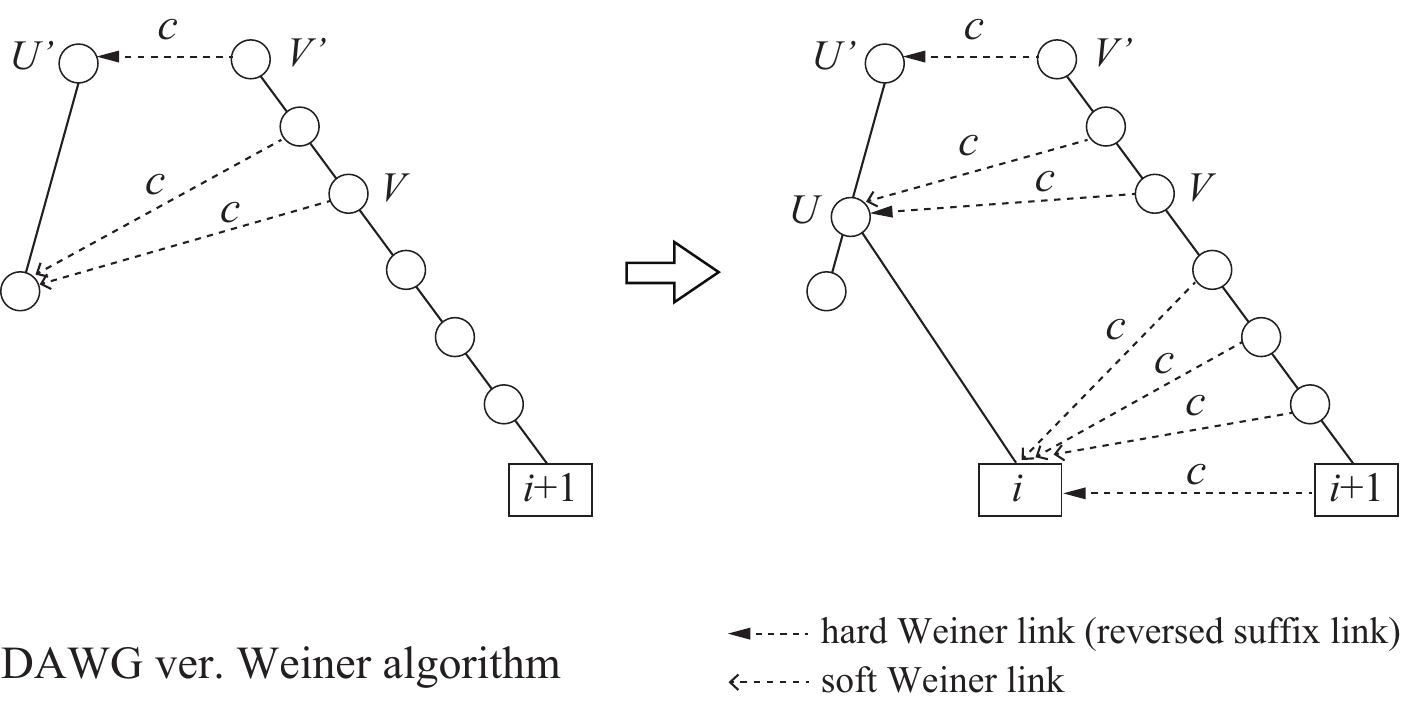}}
        \fbox{\includegraphics[scale=0.45]{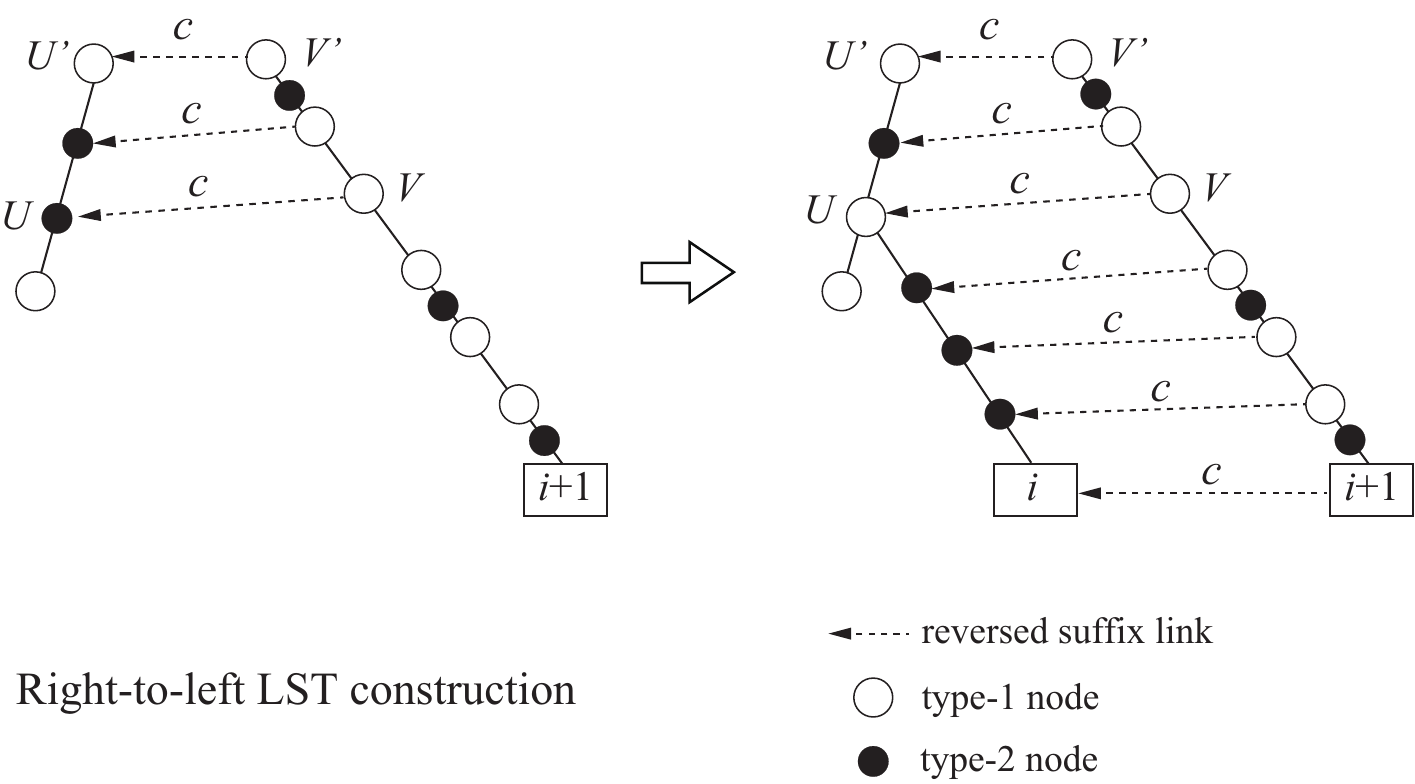}}
	\caption{
		Upper: The DAWG version of Weiner's algorithm when updating the suffix tree for $T[i+1:]$ to the suffix tree for $T[i:]$. Lower: Our right-to-left LST construction when updating $\Tree_{i+1} = \LST(T[i+1:])$ to $\Tree_{i} = \LST(T[i:])$.
	}
	\label{fig:Weiner_LST}
\end{figure}

Let us first recall Weiner's suffix tree contraction algorithm
on which our right-to-left LST construction algorithm is based.
Weiner's algorithm uses the reversed suffix links of the suffix tree
called \emph{hard Weiner links}.
In particular, we consider the version of Weiner's algorithm
that also explicitly maintains \emph{soft-Weiner links}~\cite{BreslauerI13} of the suffix tree.
In the suffix tree of a string $T$,
there is a soft-Weiner link for a node $V$ with a symbol $c$
iff $cV$ is a substring of $T$ but $cV$ is not a node in the suffix tree.
It is known that the hard-Weiner links
and the soft-Weiner links are respectively equivalent to
the primary edges and the secondary edges of the
\emph{directed acyclic word graph} (\emph{DAWG})
for the reversal of the input string~\cite{Blumer1985}.

Given the suffix tree for $T[i+1:]$,
Weiner's algorithm walks up from the leaf representing $T[i+1:]$
and first finds the nearest branching ancestor $V$ such that
$cV$ is a substring of $T[i+1:]$,
and then finds the nearest branching ancestor $V'$ such that $cV' = U'$
is also a branching node, where $c = T[i]$.
Then, Weiner's algorithm finds the insertion point for a new leaf for $T[i:]$
by following the reversed suffix link (i.e. the hard-Weiner link) from $V'$ to $U'$,
and then walking down the corresponding out-edge of $U'$ with
the difference of the string depths of $V$ and $V'$.
A new branching node $U$ is made at the insertion point
if necessary.
New soft-Weiner links are created from the nodes between the leaf for $T[i+1:]$
and $V$ to the new leaf for $T[i:]$.

Now we consider our right-to-left LST construction.
See the lower diagram of Figure~\ref{fig:Weiner_LST} for illustration.
The major difference between the DAWG version of Weiner's algorithm
and our LST construction is that
in our LST we explicitly create type-2 nodes which are 
the destinations of the soft-Weiner links.
Hence, in our linear-size suffix trie construction,
for every type-1 node between $V$ and the leaf for $T[i+1:]$,
we explicitly create a unique new type-2 node
on the path from the insertion point to the new leaf for $T[i:]$,
and connect them by the reversed suffix link labeled with $c$.
Also, we can directly access the insertion point $U$
by following the reversed suffix link of $V$,
since $U$ is already a type-2 node before the update.

The above observation also gives rise to the number of type-2 nodes in the LST.
Blumer et al.~\cite{Blumer1985} proved that
the number of secondary edges in the DAWG of any string of length $n$
is at most $n-1$.
Hence we have:
\begin{lemma}
  The number of type-2 nodes in the LST of any string of length $n$
  is at most $n-1$.
\end{lemma}
The original version of Weiner's
suffix tree construction algorithm only maintains a Boolean value
indicating whether there is a soft-Weiner link from each node with each symbol.
We also note that the number of pairs of nodes and symbols for which
the indicators are true is the same as the number of soft-Weiner links
(and hence the DAWG secondary edges).

We have seen that LSTs can be seen as a representation of Weiner's suffix trees
or the DAWGs for the reversed strings.
Another crucial point is that Weiner's algorithm only needs to read
the first symbols of edge labels.
This enables us to easily extend Weiner's suffix tree algorithm
to our right-to-left LST construction.
Below, we will give more detailed properties of LSTs and
our right-to-left construction algorithm.

Let us first observe relations between $\Tree_{i}$ and $\Tree_{i+1}$.
\begin{lemma}\label{lem:rltype2to1}
  Any non-leaf type-1 node $U$ in $\Tree_{i}$ exists in $\Tree_{i+1}$ as a type-1 or type-2 node.
\end{lemma}
\begin{proof}
  If there exist two distinct symbols $a, b \in \Sigma$
  such that $Ua, Ub$ are substrings of $T[i+1:]$,
  then clearly $U$ is a type-1 node in $\Tree_{i+1}$.
  Otherwise, then let $b$ be a unique symbol
  such that $Ub$ is a substring of $T[i+1:]$.
  This symbol $b$ exists since $U$ is not a leaf in $\Tree_{i}$.
  Also, since $U$ is a type-1 node in $\Tree_{i}$,
  there is a symbol $a \neq b$ such that
  $Ua$ is a substring of $T[i:]$.
  Note that in this case $Ua$ is a prefix of $T[i:]$
  and this is the unique occurrence of $Ua$ in $T[i:]$.
  Now, let $U' = U[2:]$.
  Then, $U'a$ is a prefix of $T[i+1:]$.
  Since $U'b$ is a substring of $T[i+1:]$,
  $U'$ is a type-1 node in $\Tree_{i+1}$
  and hence $U$ is a type-2 node in $\Tree_{i+1}$.
\end{proof}
As described above,
only a single leaf is added to the tree when updating $\Tree_{i+1}$ to $\Tree_{i}$.
The type-2 node of $\Tree_{i+1}$ that becomes type-1 in $\Tree_{i}$
is the \emph{insertion point} of this new leaf.

\begin{lemma}\label{lem:rlnewbranch}
	Let $U$ be the longest prefix of $T[i:]$
	such that $U$ is a prefix of $T[j:]$ for some $j > i$.
	Then, $U$ is a node in $\Tree_{i+1}$. 
\end{lemma}
\begin{proof}
	If $U=\varepsilon$ then $U$ is the root.
	Otherwise, since $U$ occurs twice or more in $T[i:]$ and $T[i:i+|U|] \ne T[j:j+|U|]$, 
	$U$ is a type-1 node in $\Tree_{i}$.
	 By Lemma~\ref{lem:rltype2to1}, $U$ is a node in $\Tree_{i+1}$.
\end{proof}

By \Cref{lem:rlnewbranch},
we can construct $\Tree_{i}$ by adding a branch on node $U$,
where $U$ is the longest prefix of $T[i:]$ such that $U$ is a prefix of $T[j:]$ for some $j > i$.
This node $U$ is the insertion point for $\Tree_{i}$.
The insertion point $U$ can be found by following the reversed suffix link
labeled by $c$ from the node $U[2:]$ i.e. $U = \Rlink(U[2:],c)$.
Since $U$ is the longest prefix of $T[i:]$ where $U[2:]$ occurs at least twice in $T[i+1:]$,
$U[2:]$ is the deepest ancestor of the leaf $T[i+1:]$ that has the reversed suffix link labeled by $c$.
Therefore, we can find $U$ by checking the reversed suffix links of
the ancestors of $T[i+1:]$ walking up from the leaf.
We call this leaf representing $T[i+1:]$ as the \emph{last leaf} of $\Tree_{i+1}$.

After we find the insertion point, 
we add some new nodes.
First, we consider the addition of new type-1 nodes.
\begin{lemma}\label{lem:rlnewtype1}
	There is at most one type-1 node $U$ in $\Tree_{i}$ such that 
	$U$ is a type-2 node in $\Tree_{i+1}$.
	If such a node $U$ exists, then $U$ is the insertion point of $\Tree_{i}$.
\end{lemma}
\begin{proof}
	Assume there is a type-1 node $U$ in $\Tree_{i}$ such that $U$ is a type-2 node in $\Tree_{i+1}$.
	There are suffixes $UV$ and $UW$ such that $|V| > |W|$ and $V[1] \ne W[1]$.
	Since $U$ is a type-2 node in $\Tree_{i+1}$, $UV = T[i:]$ and $UW = T[j:]$ for some $j > i$.
	Clearly, such a node is the only one which is the branching node.
\end{proof}
From Lemma~\ref{lem:rlnewtype1},
we know that new type-1 node is added at the insertion point
if it is a type-2 node.
The only other new type-1 node is the new leaf representing $T[i:]$.

\begin{figure}[t]
	\centering
	\begin{minipage}[t]{0.49\hsize}
		\centering
		\includegraphics[scale=1]{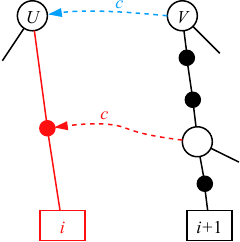}\\
		\ \ \ \small{(a)}
	\end{minipage}
	\begin{minipage}[t]{0.49\hsize}
		\centering
		\includegraphics[scale=1]{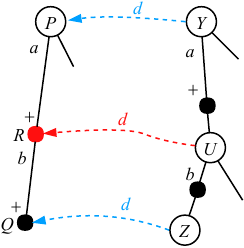}\\
		\ \ \ \small{(b)}
	\end{minipage}
	\caption{
		Illustration of (a) new branch addition and (b) type-2 nodes addition.
		The new nodes, edges, and reversed suffix links are colored red.
	}
	\label{fig:rtol_update}
\end{figure}

Next, we consider the addition of the new branch from the insertion point.
By \Cref{lem:rlnewtype1}, there are no type-1 nodes between the insertion point and
the leaf for $T[i:]$ in $\Tree_{i}$.
Thus, any node $V$ in the new branch is a type-2 node
and this node is added if $V[2:]$ is a type-1 node.
This can be checked by ascending from leaf $T[i+1:]$ to $U[2:]$,
where $U$ is the insertion point.
Regarding the labels of the new branch,
for any new node $V$ and its parent $W$,
the label of $(W,V)$ edge is the same as the label of the first edge between $W[2:]$ and $V[2:]$.
The node $V$ is a $\Plus$-node if $V[2:]$ is a $\Plus$-node or there is a node between $W[2:]$ and $V[2:]$.
\Cref{fig:rtol_update} (a) shows an illustration of the branch addition:
$V$ can be found by traversing the ancestors of $i+1$ leaf.
After we find the insertion point $U = \Rlink(V,c)$,
we add a new leaf $i$ and type-2 nodes for each type-1 node between $i+1$ leaf and $V$.

Last, consider the addition of type-2 nodes when updating the insertion point $U$ to a type-1 node.
In this case, we add a type-2 node $dU$ for any $d \in \Sigma$ such that $dU$ occurs in $T[i:]$.
\begin{lemma}\label{lem:rlnewtype2}
 Let $U$ be the insertion point of $\Tree_{i}$.
 Consider the case where $U$ is a type-2 node in $\Tree_{i+1}$.
 Let $Z$ be the nearest type-1 descendant of $U$
 and $Y$ be the nearest type-1 ancestor of $U$ in $\Tree_{i+1}$.
 For any node $Q$ such that $Q = \Rlink(Z,d)$ for some $d \in \Sigma$,
 $P = \Rlink(Y,d)$ is the parent of $Q$ in $\Tree_{i+1}$
 and there is a type-2 node $R$ between $P$ and $Q$ in $\Tree_{i}$. 
\end{lemma}
\begin{proof}
	First, we prove that $P$ is the parent of $Q$ in $\Tree_{i+1}$.
	Assume on the contrary that $P$ is not the parent of $Q$.
	Then, there is a node $Q[:j] = dZ[:j-1]$ for some $|P| < j < |Q|$.
	Thus, $Z[:j-1]$ is a type-1 ancestor of $Z$ and a type-1 descendant of $Y$,
	however, this contradicts the definition of $Z$ or $Y$.
	
	Second, we prove that there is a type-2 node between $P$ and $Q$ in $\Tree_{i}$.
	Since $U$ is a type-2 node in $\Tree_{i+1}$ and $Q = dZ$ is a node in $\Tree_{i+1}$,
	$dU$ occurs in $T[i+1:]$ but is not a node in $\Tree_{i+1}$.
	Since $U$ is a type-1 node in $\Tree_{i}$,
	$dU$ is a type-2 node $\Tree_{i}$.
\end{proof}

See \Cref{fig:rtol_update} (b) for an illustration of type-2 nodes addition.
It follows from Lemma~\ref{lem:rlnewtype2} that we can find the position of new type-2 nodes
by first following the reversed suffix link of the nearest type-1 descendant
$Z$ of $U$ in $\Tree_{i+1}$.
Then, we obtain the parent $P$ of $Q = \Rlink(Z, d)$,
and obtain $Y$ by following the suffix link of $P$.
The string depth of a new type-2 node $R$ is equal to the string depth of $U$ plus one.
We can determine whether $R$ is a $\Plus$-node
using the difference of the string depths of $Y$ and $U$.
By Lemma~\ref{lem:rltype2to1},
the total number of type-2 nodes added in this way
for all positions $1 \leq i \leq n$ is bounded
by the number of type-1 and type-2 nodes in $\Tree_{n}$ for the whole string $T$.

Algorithm~\ref{alg:rtol}
shows a pseudo-code of our right-to-left linear-size suffix trie construction algorithm.
For each symbol $c = T[i]$ read,
the algorithm finds the deepest node $V$ in the path from the root to the last leaf
for $T[i+1:]$ for which $U = \Rlink(V,c)$ is defined,
by walking up from the last leaf (line \ref{line:rtol_findbranching}).
If the insertion point $\nextNode = U = \Rlink(V,c)$ is a type-1 node,
the algorithm creates a new branch.
Otherwise (if $\nextNode$ is a type-2 node),
then the algorithm updates $\nextNode$ to type-1 and adds a new branch.
The branch addition is done in lines~\ref{line:rtol_addbranch_start}--\ref{line:rtol_addbranch_end}.

\begin{figure}[!p]
	\centering
	\includegraphics[scale=0.93]{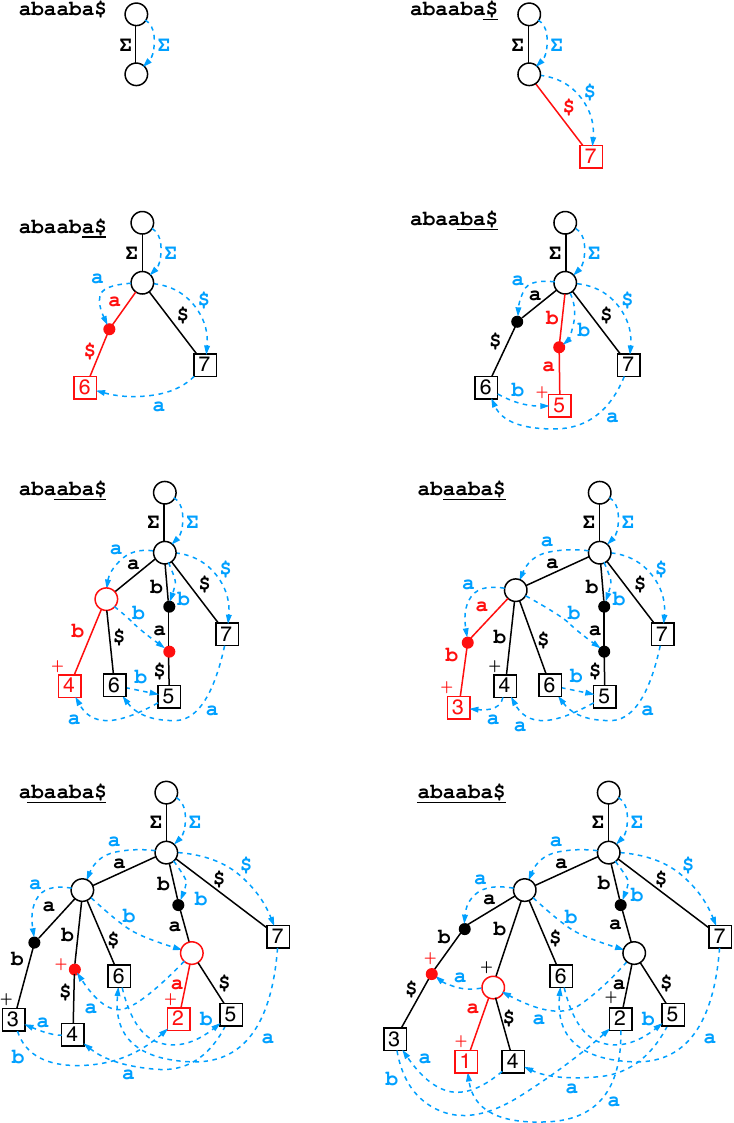}
	\vspace{3mm}
	\caption{
		A snapshot of right-to-left online construction of $\LST(T)$ with $T = \mathtt{abaaba\texttt{\$}}$ by Algorithm~\ref{alg:rtol}.
		The white circles show Type-1 nodes, the black circles show Type-2 nodes,
		and the rectangles show leaves. 
		The reversed suffix links and their labels are colored blue.
		The new branches and nodes are colored red.
	}
	
	\label{fig:rtol_example}
\end{figure}

Also, the algorithm adds nodes $R$ such that $R = \Rlink(\nextNode,d)$ for some $d \in \Sigma$ in $\Tree_{i}$.
The algorithm finds the locations of these nodes by checking the reversed suffix links of the nearest type-1 ancestor and descendant of $\nextNode$ by using $\CreateTypeTwo(\nextNode)$.
Let $Y$ be the nearest type-1 ancestor of $\nextNode$ and $Z$ be the nearest type-1 descendant of $\nextNode$ in $\Tree_{i+1}$.
For a symbol $d$ such that $\Rlink(Z,d)$ is defined,
let $P = \Rlink(Y,d)$ and $Q = \Rlink(Z,d)$:
the algorithm creates type-2 node $R$ and connects it to $P$ and $Q$.
A snapshot of right-to-left LST construction is shown in
Figure~\ref{fig:rtol_example}.

\begin{algorithm2e}[!t]
	\caption{Right-to-left linear-size suffix trie construction algorithm}
	\label{alg:rtol}
	\SetVlineSkip{0.5mm}
	$\Child(\bot,c) \ot \Root$ for any $c \in \Sigma$;
	$\Rlink(\bot,c) \ot \Root$ for all $c \in \Sigma$\;
	$\PrevInsPoint \ot \bot$;
	$\PrevLeaf \ot root$;
	$\PrevLabel \ot \NULL$\;
	\For{$i=n$ \textbf{to} $1$}{
		$c \ot T[i]$;
		$V \ot \PrevInsPoint$\; 
		\lWhile{$\Rlink(V,c) = \NULL$}{
			$V \ot \Parent(V)$}\label{line:rtol_findbranching}
		$\nextNode \ot \Rlink(V,c)$\;
		\If{$\Type(\nextNode) = 2$}{
			$\CreateTypeTwo(\nextNode)$\;
			$\Type(\nextNode) \ot 1$\; 
		}
		create a leaf $\newLeaf$\; \label{line:rtol_addbranch_start}
		$W \ot \PrevLeaf$;
		$V \ot \PrevInsPoint$;
		$Y \ot \newLeaf$\;
		\While{$\Rlink(V,c) = \NULL$}{
			create a type-2 node $X$\;
			\eIf{$V = \PrevInsPoint$}{
				$a = \PrevLabel$;}
			{
				$a = \Label(V,W)$;}
			\lIf{$\Plus(W) = \True$ \bf{or} $\Child(V,a) \ne W$}{$\Plus(Y) \ot \True$}
			$\Child(X,a) \ot Y$;
			$\Rlink(V,c) \ot X$;
			$Y \ot X$\;
			$W \ot V$\;
			\lRepeat{$\Type(V) = 1$}{$V \ot \Parent(V)$}
		}
		\eIf{$V = \bot$}{
			$a = c$;}
		{
			$a = \Label(V,W)$;}
		\lIf{$\Plus(W) = \True$ \bf{or} $\Child(V,a) \ne W$}{$\Plus(Y) \ot \True$}
		$\Child(\nextNode,a) \ot Y$\;\label{line:rtol_addbranch_end}
		$\PrevInsPoint \ot \nextNode$;
		$\PrevLeaf \ot \newLeaf$;
		$\PrevLabel \ot a$\;
	}
\end{algorithm2e}

\begin{algorithm2e}[!t]
	\caption{$\CreateTypeTwo(U)$}
	\label{alg:createtype2}
	\SetVlineSkip{0.5mm}
	\Fn{$\CreateTypeTwo(U)$}{
		$V \ot U$;
		$b = \Label(U)$;
		$Z \ot \TChild(U,b)$\;
		\For{$d$ \rm{such that} $\Rlink(Z,d) \ne \NULL$}{
			$Q \ot \Rlink(Z,d)$\;
			$P \ot \Parent(Q)$\;
			\If{$\Slink(P) \ne \NULL$}{
				$a \ot \Label(P,Q)$\;
				$Y \ot \Slink(P)$\;
				create a type-2 node $R$\;
				$\Child(P,a) \ot R$;
				$\Child(R,b) \ot Q$\;
				\If{$\Child(Y,a) \ne U$ \bf{or} $\Plus(\Child(Y,a)) = \True$}{$\Plus(R) \ot \True$}
				\If{$\Child(U,b) \ne Z$ \bf{or} $\Plus(\Child(U,b)) = \True$}{$\Plus(Q) \ot \True$}
			}
		}
	}
	
\end{algorithm2e}

%% file: doc/rtolcomp.tex
We discuss the time complexity of our right-to-left online LST construction algorithm.
Basically, the analysis follows the amortization argument
for Weiner's suffix tree construction algorithm.
First, consider the cost of finding the insertion point for each $i$.
\begin{lemma}\label{lem:findbranchtime}
	Our algorithm finds the insertion point of $\Tree_{i}$ in $O(\log \sigma)$ amortized time.
\end{lemma}
\begin{proof}
	For each iteration, the number of type-1 and type-2 nodes we visit
        from the last leaf to find the insertion point
        is at most $\Depth(L_{i+1})-\Depth(U_i)+1$,
        where $L_{i+1}$ is the leaf representing $T[i+1:]$
        and $U_i$ is the insertion point for the new leaf representing $T[i:]$
        in $\Tree_{i}$, respectively,
        and $\Depth(X)$ denotes the depth of any node $X$ in $\Tree_{i}$.
        See also the lower diagram of Figure~\ref{fig:Weiner_LST} for illustration.
	Therefore, the total number of nodes visited
        is $\sum_{1\le i < n} \Depth(L_{i+1})-\Depth(U_i)+1 \le 2n$.
	Since finding each reversed suffix link takes $O(\log \sigma)$ time,
        the total cost for finding the insertion points for all $1 \leq i \leq n$
        is $O(n \log \sigma)$, which is amortized to $O(\log \sigma)$ per iteration.
\end{proof}

Last, the computation time of a new branch addition in each iteration is as follows.
\begin{lemma}\label{lem:addbranchtime}
  Our algorithm adds a new leaf and new type-2 nodes
  between the insertion point and the new leaf in
  $\Tree_{i}$ in $O(\log \sigma)$ amortized time.
\end{lemma}
\begin{proof}
  Given the insertion point for $\Tree_{i}$,
  it is clear that we can insert a new leaf in $O(\log \sigma)$ time.
  For each new type-2 node in the path from the insertion point and the new leaf for $T[i:]$,
  there is a corresponding type-1 node in the path above the last leaf $T[i+1:]$
  (see also the lower diagram of Figure~\ref{fig:Weiner_LST}).
  Thus the cost for inserting all type-2 nodes can be charged to
  the cost for finding the insertion point for $\Tree_{i}$,
  which is amortized $O(\log \sigma)$ per a new type-2 node by Lemma~\ref{lem:findbranchtime}.  
\end{proof}

Next, we discuss how to maintain fast links in the growing tree.
Let $U$ be the insertion point and $V$ be the new leaf in of $\Tree_{i}$.
All new nodes between $U$ and $V$ are leaves of $\SLT(\Tree_{i})$.
Thus, we can add those nodes to the edge link trees by $\AddLeaf$ while marking necessary nodes.
Moreover, if $U$ is a type-2 node in $\Tree_{i+1}$,
we perform $\AddLeaf(U)$ for each node $\Rlink(U,d)$ for some $d \in \Sigma$ and
$\Demote(W)$, where $W$ is the type-1 node that is a child of $U$ in $\Tree_{i+1}$.

By Lemmas~\ref{lem:nmainc}, \ref{lem:findbranchtime} and~\ref{lem:addbranchtime},
we get the following theorem:
\begin{theorem}\label{theorem:rtollstconstructiontime}
	Given a string $T$ of length $n$, our algorithm constructs $\LST(T)$ and its $\FLink$ in $O(n \log \sigma)$ time and $O(n)$ space online, by reading $T$ from the right to the left.
\end{theorem}

%% file: doc/ltor.tex
\section{Left-to-right online algorithm} \label{sec:left-to-right}

\begin{figure}[t]
	\centering
	\includegraphics[scale=1]{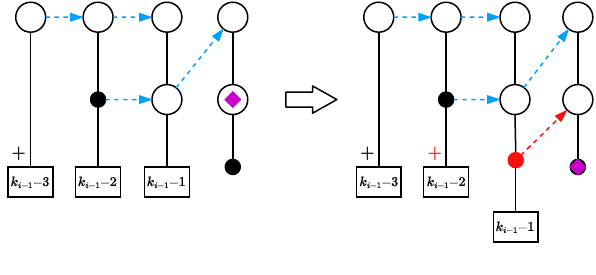}
	\caption{
		Illustration for updating the parts of $\PTree_{i-1}$ that correspond to $T[j:i-1]$ for $j < k_{i-1}$.
		The purple diamond shows the active point.
		The new $\Plus$ sign, node, and its suffix link are colored red.
	}
	\label{fig:ltor_updateleft}
\end{figure}

In this section, we present an algorithm that constructs the linear-size suffix trie of a string $T$
by reading the symbols of $T$ from the left to the right.
Our algorithm constructs a slightly-modified
data structure called the pre-LST defined as follows:
	The pre-LST $\PLST(T)$ of a string $T$ is a subgraph of $\STrie(T)$ consisting of two types of nodes,
	\begin{enumerate}
		\item Type-1: The root, branching nodes, and leaves of $\STrie(T)$.
		\item Type-2: The nodes of $\STrie(T)$ that are not type-1 nodes and their suffix links point to type-1 nodes.
	\end{enumerate}
The main difference between $\PLST(T)$ and $\LST(T)$ is the definition of type-1 nodes.
While $\LST(T)$ may contain non-branching type-1 nodes that correspond to 
non-branching internal nodes of $\STree(T)$ which represent repeating suffixes,
$\PLST(T)$ does not contain such type-1 nodes.
When $T$ ends with a unique terminal symbol $\$$,
the pre-LST and LST of $T$ coincide.

Our algorithm is based on Ukkonen's suffix tree construction algorithm~\cite{Ukkonen1995}.
For each prefix $T[:i]$ of $T$,
there is a unique position $k_i$ in $T[:i]$
such that $T[k_i:i]$ occurs twice or more in $T[:i]$
but $T[k_i-1:i]$ occurs exactly once in $T[:i]$.
In other words, $T[k_i-1:i]$ is the shortest suffix of $T[:i]$ that is represented
as a leaf in the current pre-LST $\PLST(T[:i])$,
and $T[k_i:i]$ is the longest suffix of $T[:i]$ that is
represented in the ``inside'' of $\PLST(T[:i])$.
The location of $\PLST(T[:i])$ representing the longest repeating suffix $T[k_i:i]$
of $T[:i]$ is called the \emph{active point}, as in the
Ukkonen's suffix tree construction algorithm.
We also call $k_i$ the \emph{active position} for $T[:i]$.
Our algorithm keeps track of the location of the active point (and the active position)
each time a new symbol $T[i]$ is read for increasing $i = 1, \ldots, n$.
We will show later that the active point can be
maintained in $O(\log \sigma)$ amortized time per iteration,
using a similar technique to our pattern matching algorithm on LSTs
in Lemma~\ref{lem:patmatch}.
In order to ``neglect'' extending the leaves that already exist in the current tree,
Ukkonen's suffix tree construction algorithm uses
the idea of \emph{open leaves} that do not explicitly maintain
the lengths of incoming edge labels of the leaves.
However, we cannot adapt this open leaves technique to construct pre-LST directly,
since we need to add type-2 nodes on the incoming edges of some leaves.
Fortunately, there is a nice property on the pre-LST so we can update it efficiently.
We will discuss the detail of this property later.
Below, we will give more detailed properties of pre-LSTs and
our left-to-right construction algorithm.

Let $\PTree_{i} = \PLST(T[:i])$ be the pre-LST of $T[:i]$.
Our algorithm constructs $\PTree_{i}$ from $\PTree_{i-1}$ incrementally when a new symbol $c = T[i]$ is read.

There are two kinds of leaves in $\PLST(T[:i])$,
the ones that are $\Plus$-nodes and the other ones that are not $\Plus$-nodes.
There is a boundary in the suffix link chain of the leaves
that divides the leaves into the two groups, as follows:
\begin{lemma}\label{lem:leafpoint}
	Let $T[j:i]$ be a leaf of $\PTree_{i}$, for $1 \le j < k_i$.
	There is a position $l$ such that $T[j:i]$ is a $\Plus$-node for $1 \le j < l$
	and not a $\Plus$-node for $l \le j < k_i$.
\end{lemma}
\begin{proof}
	Assume on the contrary there is a position $j$ such that $T[j:i]$ is not a $\Plus$-node
	and $T[j+1:i]$ is a $\Plus$ node.
	Since $T[j:i]$ is not a $\Plus$-node, $T[j:i-1]$ is a node.
	By definition, $T[j+1:i-1]$ is also a node.
	Thus $T[j+1:i]$ is not a $\Plus$-node, which is a contradiction.
\end{proof}
Intuitively, the leaves that are $\Plus$-nodes in $\PTree_{i}$
are the ones that were created in the last step of the algorithm with the last read symbol $T[i]$.

\begin{figure}[t]
	\centering
	\includegraphics[scale=0.8]{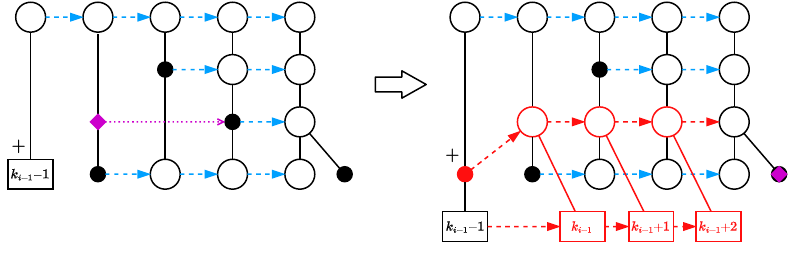}
	\caption{
		Illustration for updating the parts of $\PTree_{i-1}$ that correspond to $T[j:i-1]$ for $j \ge k_{i-1}$.
		The purple diamond and arrow show the active point and its virtual position when reading the edge.
		The new branches, nodes, and their suffix links are colored red.
	}
	\label{fig:ltor_updateright}
\end{figure}

When updating $\PTree_{i-1}$ into $\PTree_{i}$,
the active position $k_{i-1}$ for $T[:i-1]$ divides the suffixes $T[j:i-1]$ into two parts,
the $j < k_{i-1}$ part and the $j \ge k_{i-1}$ part.
First, we consider updating the parts of $\PTree_{i-1}$ that correspond to $T[j:i-1]$ for $j < k_{i-1}$.
\begin{lemma}\label{lem:lropenleaf}
	For any leaf $T[j:i-1]$ of $\PTree_{i-1}$ with $j < k_{i-1} - 1$,
        $T[j:i-1]$ is implicit in $\PTree_{i}$.
\end{lemma}
\begin{proof}
	Consider updating $\PTree_{i-1}$ to $\PTree_{i}$.
	$T[k_{i-1}-1:i-1]$ cannot be a type-1 node in $\PTree_{i}$.
	Therefore, $T[k_{i-1}-2:i-1]$ is implicit in $\PTree_{i}$.
	Similarly, $T[j:i-1]$ is implicit in $\PTree_{i}$, for any $j < k_{i-1}-1$.
\end{proof}

\begin{lemma}\label{lem:lrplusleaf}
  If $T[j:i-1]$ is a leaf in $\PTree_{i-1}$,
  then $T[j:i]$ is a $\Plus$-leaf in $\PTree_{i}$, where $1 \le j < k_{i-1}-1$.
\end{lemma}
\begin{proof}
  Assume on the contrary that $T[j:i-1]$ is a leaf in $\PTree_{i-1}$
  but $T[j:i]$ is not a $\Plus$-leaf in $\PTree_{i}$.
  Then $T[j:i-1]$ is a node in $\PTree_{i}$.
  Since $T[j:i-1]$ is a leaf in $\PTree_{i-1}$,
  $T[j:i-1]$ cannot be a type-1 node in  $\PTree_{i}$.
  Moreover, $T[j+1:i-1]$ is a leaf in $\PTree_{i-1}$,
  thus $T[j+1:i-1]$ cannot be a type-1 node in $\PTree_{i}$ and $T[j:i-1]$ cannot be a type-2 node in $\PTree_{i}$.
  Therefore, $T[j:i-1]$ is neither type-1 nor type-2 node in $\PTree_{i}$, which contradicts the assumption.
\end{proof}
\Cref{lem:lropenleaf} shows that we do not need to add nodes on the leaves of $\mathcal{P}_{i-1}$ besides $T[k_{i-1}-1:i-1]$ leaf and \Cref{lem:lrplusleaf} shows that we can update all leaves
$T[j:i-1]$ for $1 \le j < k_{i-1}-1$ to a $\Plus$-leaf.
Therefore, besides the leaf for $T[k_{i-1}-1:i-1]$, once we update a leaf to $\Plus$ node, 
we do not need to update it again.
\Cref{fig:ltor_updateleft} shows an illustration of how to update this part.
Lines~\ref{line:plus_leaf1_start}-\ref{line:plus_leaf1_end} and \ref{line:plus_leaf2_start}-\ref{line:plus_leaf2_end} 
of Algorithm~\ref{alg:ltor} show the procedure to update the leaves.

Next, we consider updating non-leaf nodes of $\PTree_{i-1}$ to $\PTree_{i}$.
Similarly to the Ukkonen's suffix tree construction algorithm,
we can see that any type-1 non-leaf node of $\PTree_{i-1}$ is also a type-1 node in $\PTree_{i}$. 
However, implicit and type-2 nodes of $\PTree_{i-1}$ could be a different type of node in $\PTree_{i}$.
Fortunately, we do not need to update the nodes that did not corresponded to a suffix of $T[:i-1]$.
\begin{lemma}
	Let $U$ be a type-2 node or implicit node of $\PTree_{i-1}$ that is not a suffix of $T[:i-1]$.
	Then, $U$ is a type-2 node or implicit node of $\PTree_{i}$,
	respectively.
\end{lemma}
\begin{proof}
	Let $U$ be a type-2 node of $\PTree_{i-1}$ that is not a suffix of $T[:i-1]$.
	By the definition of type-2 node,
	there exist symbols $a$, $b$, and $c$ such that  
	$a \ne b$,
	both $U[2:]a$ and $U[2:]b$ occur in $T[:i-1]$,
	$Uc$ occurs in $T[:i-1]$,
	and $Ud$ does not occur in $T[:i-1]$ for any symbol $d \ne c$.
	Note that $c$ can be equal to $a$ or $b$.
	Since $U$ is not a suffix of $T[:i-1]$,
	$Ud$ does not occur in $T[:i]$ for any symbol $d \ne c$.
	Therefore, $U$ is a type-2 node of $\PTree_{i}$.
	The case $U$ is an implicit node can be proved in a similar way.
\end{proof}

\begin{figure}[t]
	\centering
	\includegraphics[scale=0.5]{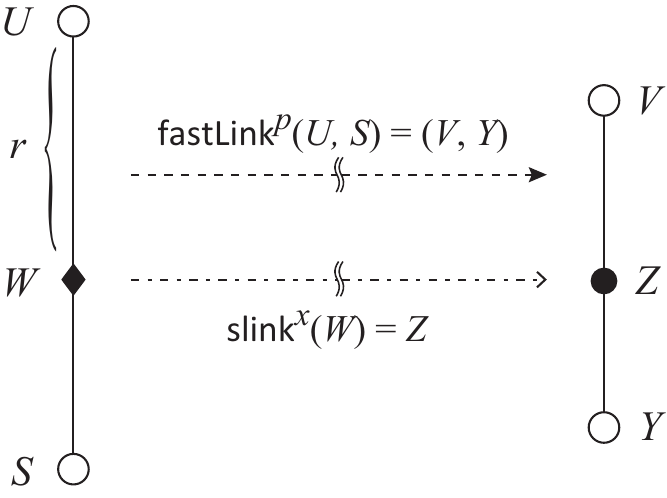}
	\caption{Illustration for our analysis of the cost to maintain the active point. The diamond shows the current location of the active point. New leaves will be created from $W$ to $Z$ by following the (virtual) suffix link chain of length $x$. When we have reached the edge $(V, Y)$, we have already retrieved the corresponding prefix of the label between $U$ and $W$. The rest of the label can be retrieved by at most $r$ applications of $\FLink$ from edge $(V, Z)$.}
	\label{fig:active_point}
\end{figure}

On the other hand, we need to consider updating the parts of $\PTree_{i-1}$ that correspond to $T[j:i-1]$ for $j \ge k_{i-1}$.
If $T[k_{i-1}:i]$ exists in the current pre-LST (namely $T[k_{i-1}:i]$ occurs in $T[:i-1]$),
we do not need to update the parts of $\PTree_{i-1}$ that correspond to suffixes $T[j:i-1]$ for $j \ge k_{i-1}$.
Then we have $k_{i} = k_{i-1}$ and $T[k_{i}:i]$ is the active point of $\PTree_{i}$.
Otherwise, we need to create new nodes recursively from the active point
that will be the parent of each new leaf.
There are three cases for the active point $T[k_{i-1}:i-1]$ in $\PTree_{i-1}$:

\textbf{Case 1}: $T[k_{i-1}:i-1]$ is a type-1 node in $\PTree_{i-1}$.
Let $T[p:i]$ be the longest suffix of $T[k_{i-1}:i]$ that exists in $\PTree_{i-1}$.
Since $T[k_{i-1}:i-1]$ is a type-1 node, $T[j:i-1]$ is also a type-1 node for $k_{i-1} \le j < p$.
Therefore, we can obtain $\PTree_{i}$ by adding a leaf from the node representing
$T[j:i-1]$ for every $k_{i-1} \le j < p$,
with edge label $c$ by following the suffix link chain from $T[k_{i-1}:i-1]$.
Then, we add one new type-2 node, which is $T[k_{i-1}-1:i-1]$
that is connected to the type-1 node $T[k_{i-1}:i-1]$ by the suffix link.
Moreover, $p$ will be the active position for $T[:i]$, namely $k_{i} = p$.

\textbf{Case 2}: $T[k_{i-1}:i-1]$ is a type-2 node in $\PTree_{i-1}$.
Similarly to Case 1,
we add a leaf from the node representing $T[j:i-1]$ for every $k_{i-1} \le j < p$
with edge label $c$ by following the suffix link chain from $T[k_{i-1}:i-1]$,
where $p$ is defined as in Case 1.
Then, $T[k_{i-1}:i-1]$ becomes a type-1 node,
and a new type-2 node $T[k_{i-1}-1:i-1]$ is added and is connected to
this type-1 node $T[k_{i-1}:i-1]$ by the suffix link.
Moreover, for any symbol $d$ such that $dT[k_{i-1}:i-1]$ is a substring
of $T[:i]$, a new type-2 node for $dT[k_{i-1}:i-1]$ is added to the tree,
and is connected by the suffix link to this new type-1 node $T[k_{i-1}:i-1]$.
These new type-2 nodes can be found in the same way as
in Lemma~\ref{lem:rlnewtype2} for our right-to-left LST construction.
Finally, $p$ will become the active position for $T[:i]$, namely $k_i = p$.

\textbf{Case 3}: $T[k_{i-1}:i-1]$ is implicit in $\PTree_{i-1}$.
In this case, there is a position $p > k_{i-1}$ such that $T[p:i-1]$ is a type-2 node.
We create new type-1 nodes $T[j:i-1]$ and leaves $T[j:i]$ for $k_{i-1} \le j < p$,
then do the same procedure as Case 2 for $T[j:i-1]$ for $p \le j$.

\Cref{fig:ltor_updateright} shows an illustration of how to add new leaves.
Algorithm~\ref{alg:ltor} shows a pseudo-code of our 
left-to-right online algorithm for constructing LSTs.
Here, $\Leaf[i]$ denotes the leaf that corresponded to the $i$-th longest suffix $T[i:]$.
In Case 1 or Case 2, the algorithm
checks whether there is an outgoing edge labeled with $c = T[i]$,
and performs the above procedures 
(lines~\ref{line:ltor_addbranch_start}--\ref{line:ltor_addbranch_end}).
In Case 3, we perform $\ReadEdge$ to check if the active point can
proceed with $c$ on the edge.
The function $\ReadEdge$ returns the location of the new active point and sets $\Flag=\False$ if there is no mismatch, or it returns the mismatching position and sets $\Flag=\True$
if there is a mismatch.
If there is no mismatch, then we just update the $T[j:i-1]$ part of the current LST
for $j < k_{i-1}$.
Otherwise, then we create new nodes as explained in Case 3,
by $\Split$ in the pseudo-code.
A snapshot of left-to-right LST construction is shown in
Figure~\ref{fig:ltor_example}.

However, there is a special case for $\FLink(U,V)$ when $V$ is the leaf with the largest label, namely $\Leaf[k-1]$,
because $V$ does not have suffix links.
In the case $U = \activeNode$,
the path label of $(U,V)$ is equivalent to the path label of $(\Slink(U),U)$,
thus we can simulate $\FLink(U,V)$ by $(\Slink(U),U)$ if there exists a node between $\Slink(U)$ and $U$,
or by $\FLink(\Slink(U),U)$ otherwise.
In the case $U \ne \activeNode$,
that is when we need to read $(U,V)$ after some recursion from $\activeNode$,
we have the active point on an edge between a leaf $W$ and its parent.
In this case, since both edges connected with leaves $V$ and $W$ will grow,
we can simulate $\FLink(U,V)$ by $(\Slink(U),W)$ if there exists a node between $\Slink(U)$ and $U$,
or by $\FLink(\Slink(U),W)$ otherwise.
Note that we do not show this procedure explicitly in Algorithm~$\ref{alg:readedge}$ for simplicity.

%% file: doc/ltorcomp.tex
\begin{algorithm2e}[!t]
	\caption{Left-to-right linear-size suffix trie construction algorithm}
	\label{alg:ltor}
	\SetVlineSkip{0.5mm}
	create $\Root$ and $\bot$;
	$\Child(\bot,c) \ot \Root$ for any $c \in \Sigma$\;
	$\activeNode = \Root$;
	$i \ot 1$;
	$l \ot 1$;
	$k \ot 1$\;
	\While{$i \le n$}{
		$c \ot T[i]$\;
		\If{$\Child(\activeNode,c) \ne \NULL$}{
			$V \ot \Child(\activeNode,c)$\;
			$(U,i',\Flag) \ot \ReadEdge((\activeNode,V),i)$\;
			\eIf{$\Type(\activeNode) = 1$}{
				create a type-2 node $W$\;
				$V \ot \Parent(\Leaf[k-1])$\;
				$\Child(W,c) \ot \Leaf[k-1]$;
				$\Child(V,\Label(V,\Leaf[k-1])) \ot W$\;
				$\Plus(W,c) \ot \Plus(\Leaf[k-1])$;
				$\Slink(W) \ot \activeNode$;
			}
			{
				$\Plus(\Leaf[k-1]) \ot \True$;}
			\While{$l \ne k-1$}{\label{line:plus_leaf1_start}
				$\Plus(\Leaf[l]) \ot \True$\;
				$l \ot l + 1$\;}\label{line:plus_leaf1_end}
			\eIf{$\Flag = \False$}{
				\lIf{$\Plus(U) = \True$}{
					$\Plus(\Leaf[k-1]) \ot \True$}
			}
			{
				$\Split(U,\activeNode,c,i,i')$;}
			$\activeNode \ot U$;	
			$i \ot i'$\;
		}
		\Else{\label{line:ltor_addbranch_start}			
			\If{$\Type(\activeNode) = 2$}{
				$\CreateTypeTwo(\activeNode)$;
				$\Type(\activeNode) \ot 1$\;
			}
			\While{$l \ne k-1$}{\label{line:plus_leaf2_start}
				$\Plus(\Leaf[l]) \ot \True$\;
				$l \ot l + 1$\;}\label{line:plus_leaf2_end}
			create a type-2 node $W$;
			$V \ot \Parent(\Leaf[k-1])$\;
			$\Child(W,c) \ot \Leaf[k-1]$;
			$\Child(V,\Label(V,\Leaf[k-1])) \ot W$\;
			$\Plus(W,c) \ot \Plus(\Leaf[k-1])$;
			$\Slink(W) \ot \activeNode$\;
			\While{$\Child(\activeNode,c) = \NULL$}{
				create a leaf $U$\;
				$\Child(\activeNode,c) \ot U$;
				$\Slink(\Leaf[k-1]) \ot U$\;
				$k \ot k + 1$;
				$\Leaf[k-1] \ot U$;
				$\activeNode = \Slink(\activeNode)$\;
			}
		}\label{line:ltor_addbranch_end}
	}
\end{algorithm2e}

\begin{algorithm2e}[!t]
	\caption{$\ReadEdge((U,V),i)$}
	\label{alg:readedge}
	\SetVlineSkip{0.5mm}
	\Fn{$\ReadEdge(U,V,i)$}{
		\While{$U \ne V$}{
			$c \ot T[i]$\;
			\lIf{$\Child(U,c) = \NULL$}{
				$\Ret$ $(U,i,\True)$}
			\Else{
				\If{$\Plus(\Child(U,c)) = \True$}{
					$(W,i,\Flag) \ot \ReadEdge(\FLink(U,\Child(U,c)),i)$\;
					\lIf{$\Flag = true$}{
						$\Ret$ $(W,i,\True)$}
					$U \ot W$\;
				}
				\lElse{
					$U \ot \Child(U,c)$;
					$i \ot i + 1$}
			}
		}
		$\Ret$ $(U,i,\False)$\;
	}
	
\end{algorithm2e}

\begin{algorithm2e}[!t]
	\caption{$\Split(U,X,a,i,i')$}
	\label{alg:split}
	\SetVlineSkip{0.5mm}
	\Fn{$\Split(U,X,a,i,i')$}{
		$b = \Label(U,\Child(U))$;
		$c' \ot T[i']$\;
		create a type-2 node $W$\;
		$V \ot \Parent(\Leaf[k-1])$\;
		$\Child(W,c) \ot \Leaf[k-1]$;
		$\Child(V,\Label(V,\Leaf[k-1])) \ot W$\;
		$\Plus(W) \ot \Plus(\Leaf[k-1])$;
		$\LastNode \ot W$\;
		$k \ot k+1$;
		$Y' \ot \Leaf[k-1]$\;
		\While{$X \ne U$}{
			\eIf{$\Type(X) = 1$}{
				$Y \ot \Child(X,a)$\;}
			{
				$Y \ot \Child(X)$\;
			}
			$d \ot \STriedepth(Y) - \STriedepth(X)$\;
			\While{$d < i' - i$}{
				$X \ot Y$;
				$i \ot i + d$\;
				$Y \ot \Child(X)$;	
				$d \ot \STriedepth(Y) - \STriedepth(X)$\;
			}
			\If{$X \ne U$}{
				create a type-1 node $Z$;
				create a leaf $Y'$;
				$a \ot \Label(X,Y)$\;
				$\Child(X, a) \ot Z$;
				$\Child(Z, b) \ot Y$;
				$\CreateTypeTwo(Z)$\;
				$\Child(Z, c') \ot Y'$\;
				\lIf{$i' - 1 > 1$}{$\Plus(Z) \ot \True$}
				\lIf{$d - (i' - 1) > 1$}{$\Plus(Y) \ot \True$}
				$\Slink(\LastNode) \ot Z$;
				$\Slink(\Leaf[k-1]) \ot Y'$\;
				$k \ot k + 1$;
				$\Leaf[k-1] \ot Y'$\;
				$\LastNode \ot Z$;
				$X \ot \Slink(X)$\;
			}
		}
		$\Slink(\LastNode) \ot U$\;
	}
\end{algorithm2e}

\begin{figure}[!p]
	\centering
	\includegraphics[scale=0.93]{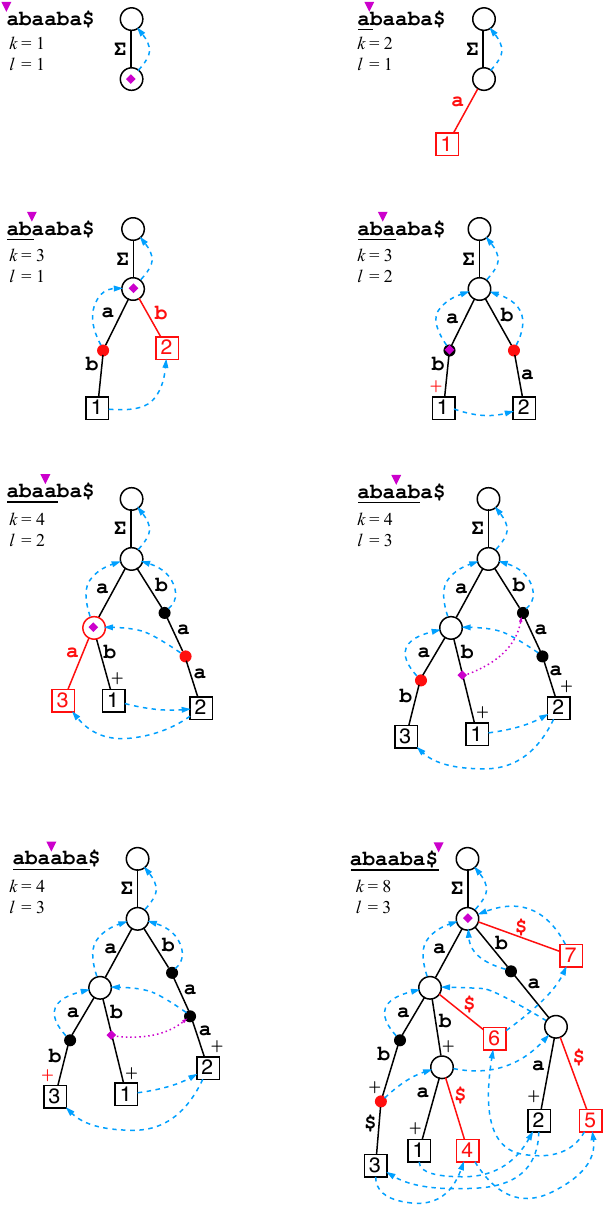}
	\vspace{3mm}
	\caption{
		A snapshot of left-to-right online construction of $\LST(T)$ with $T = \mathtt{abaaba\texttt{\$}}$ by Algorithm~\ref{alg:ltor}.
		The purple diamond and arrow represent the active point and its virtual position when reading the edge label.
		The suffix links are colored blue.
		The new branches and nodes are colored red.
		Each $k$ is the active position and $l$ is the boundary position
		for $\Plus$-leaves and non-$\Plus$ leaves defined in Lemma~\ref{lem:leafpoint}.
	}
	
	\label{fig:ltor_example}
\end{figure}

We discuss the time complexity of our left-to-right online construction for LSTs.
To maintain the active point for each $T[:i]$,
we use a similar technique to Lemma~\ref{lem:patmatch}.
\begin{lemma} \label{lem:active_point_maintain}
The active point can be maintained in $O(f(n)+\log \sigma)$ amortized time
per each iteration,
where $f(n)$ denotes the time for accessing $\FLink$ in our growing LST.
\end{lemma}

\begin{proof}
We consider the most involved case where 
the active point lies on an implicit node $W$
on some edge $(U, S)$ in the current LST.
The other cases are easier to show.
Let $r = |W|-|U|$, i.e., the active point is hanging off $U$ with string depth $r$.
Let $Z$ be the type-2 node from which a new leaf will be created.
By the monotonicity on the suffix link chain
there always exists such a type-2 node.
See Figure~\ref{fig:active_point} for illustration.
Let $p$ be the number of applications of $\FLink$ from edge $(U, S)$
until reaching the edge $(V, Y)$ on which $Z$ lies.
Since such a type-2 node $Z$ always exists,
we can sequentially retrieve the first $r$ symbols with at most $r$ applications of $\FLink$
by the same argument to Lemma~\ref{lem:patmatch}.
Thus the number of applications of $\FLink$
until finding the next location of the active point is bounded by $p+r$.
If $x$ is the number of (virtual) suffix links from $W$ to $Z$,
then $p \leq x$ holds.
Recall that we create at least $x+1$ new leaves by following the (virtual) suffix link chain
from $W$ to $Z$.
Now $r$ is charged to the number of text symbols read on the edge from $U$,
and $p$ is charged to the number of newly created leaves,
and both of them are amortized constant as in Ukkonen's suffix tree algorithm.
Thus the number of applications of $\FLink$ is amortized constant,
which implies that it takes $O(f(n)+\log \sigma)$ amortized time to maintain the active point.
\end{proof}

To maintain $\FLink$ in our growing (suffix link) tree,
we use the nearest marked ancestor (NMA) data structure that we described in \Cref{sec:ELT}.
By \Cref{lem:nmainc} we know that in order to maintain the edge link tree enhanced with the NMA data structure we need $O(1)$ amortized time,
thus the function involved in \Cref{lem:active_point_maintain} is $f(n) = O(1)$.
This leads to the final result of this section.

\begin{theorem}\label{theorem:ltorlstconstructiontime}
	Given a string $T$ of length $n$, our algorithm constructs $\LST(T)$ in $O(n \log \sigma) $ time and $O(n)$ space online, by reading $T$ from the left to the right.
\end{theorem}

%% file: doc/conclusion.tex
\section{Conclusions and Future Work} \label{sec:conclusions}

In this paper, we proposed two online construction algorithms for linear-size suffix trees (LSTs), one in a right-to-left fashion, and the other in a left-to-right fashion, both running in $O(n \log \sigma)$ time with working $O(n)$ space,
for an input string of length $n$ over an ordered alphabet of size $\sigma$.
The previous offline construction algorithm by Crochemore et al.~\cite{Crochemore2016} needs to construct suffix trees as intermediate structures,
in which each edge label is a pair of positions in the input string.
As a consequence, their algorithm requires storing the input string to construct the suffix tree, although the input string can be deleted after the LST is constructed.
On the other hand, our algorithms construct the LST directly
without constructing suffix trees as intermediate structures.
This allows our methods not to store the input string when constructing LSTs.

Fischer and Gawrychowski~\cite{0001G15} showed
how to build suffix trees in a right-to-left online manner
in $O(n(\log\log n + \log^2 \log \sigma / \log \log \log \sigma))$ time
for an integer alphabet of size $\sigma = n^{O(1)}$.
It might be possible to extend their result to 
our right-to-left online LST construction algorithm.

Takagi et al.~\cite{Takagi2017} proposed \emph{linear-size CDAWGs} (\emph{LCDAWG}),
which are edge-labeled DAGs obtained by merging isomorphic subtrees of LSTs.
They showed that the LCDAWG of a string $T$ takes only $O(e+e')$ space,
where $e$ and $e'$ are respectively the numbers of right and left extensions of
the maximal repeats in $T$,
which are always smaller than the text length $n$.
Belazzougui and Cunial~\cite{BelazzouguiC17} proposed
a very similar CDAWG-based data structure that uses only $O(e)$ space.
It is not known whether these data structures can be efficiently constructed in an online manner,
and thus it is interesting to see if our algorithms can be extended to these data structures.
The key idea to both of the above CDAWG-based structures is to implement
edge labels by \emph{grammar-compression} or \emph{straight-line programs},
which are enhanced with efficient grammar-compressed data structures~\cite{GasieniecKPS05,BilleLRSSW15}.
In our online setting, the underlying grammar needs to be dynamically updated,
but these data structures are static.
It is worth considering if these data structures can be efficiently dynamized
by using recent techniques such as e.g.~\cite{GawrychowskiKKL18}.

%% file: arxiv.bbl
\begin{thebibliography}{10}

\bibitem{Alstrup1998}
Stephen Alstrup, Thore Husfeldt, and Theis Rauhe.
\newblock Marked ancestor problems.
\newblock In {\em Proceedings of the 39th Annual Symposium on Foundations of
  Computer Science}, pages 534--544, 1998.
\newblock \href {https://doi.org/10.1109/SFCS.1998.743504}
  {\path{doi:10.1109/SFCS.1998.743504}}.

\bibitem{BelazzouguiC17}
Djamal Belazzougui and Fabio Cunial.
\newblock Fast label extraction in the {CDAWG}.
\newblock In {\em Proceedings of the 24th International Symposium on String
  Processing and Information Retrieval}, pages 161--175, 2017.
\newblock \href {https://doi.org/10.1007/978-3-319-67428-5\_14}
  {\path{doi:10.1007/978-3-319-67428-5\_14}}.

\bibitem{BilleLRSSW15}
Philip Bille, Gad~M. Landau, Rajeev Raman, Kunihiko Sadakane, Srinivasa~Rao
  Satti, and Oren Weimann.
\newblock Random access to grammar-compressed strings and trees.
\newblock {\em {SIAM} Journal on Computing}, 44(3):513--539, 2015.
\newblock \href {https://doi.org/10.1137/130936889}
  {\path{doi:10.1137/130936889}}.

\bibitem{Blumer1985}
Anselm Blumer, J.~Blumer, David Haussler, Andrzej Ehrenfeucht, M.T. Chen, and
  Joel Seiferas.
\newblock {The smallest automation recognizing the subwords of a text}.
\newblock {\em Theoretical Computer Science}, 40:31--55, 1985.
\newblock \href {https://doi.org/10.1016/0304-3975(85)90157-4}
  {\path{doi:10.1016/0304-3975(85)90157-4}}.

\bibitem{Blumer1987}
Anselm Blumer, J.~Blumer, David Haussler, Ross McConnell, and Andrzej
  Ehrenfeucht.
\newblock {Complete inverted files for efficient text retrieval and analysis}.
\newblock {\em Journal of the ACM}, 34(3):578--595, 1987.
\newblock \href {https://doi.org/10.1145/28869.28873}
  {\path{doi:10.1145/28869.28873}}.

\bibitem{BreslauerI13}
Dany Breslauer and Giuseppe~F. Italiano.
\newblock Near real-time suffix tree construction via the fringe marked
  ancestor problem.
\newblock {\em Journal of Discrete Algorithms}, 18:32--48, 2013.
\newblock \href {https://doi.org/10.1016/j.jda.2012.07.003}
  {\path{doi:10.1016/j.jda.2012.07.003}}.

\bibitem{Crochemore2016}
Maxime Crochemore, Chiara Epifanio, Roberto Grossi, and Filippo Mignosi.
\newblock {Linear-size suffix tries}.
\newblock {\em Theoretical Computer Science}, 638:171--178, 2016.
\newblock \href {https://doi.org/10.1016/j.tcs.2016.04.002}
  {\path{doi:10.1016/j.tcs.2016.04.002}}.

\bibitem{Crochemore1997const}
Maxime Crochemore and Renaud V{\'{e}}rin.
\newblock {Direct construction of compact directed acyclic word graphs}.
\newblock In {\em Proceedings of 8th Annual Symposium on Combinatorial Pattern
  Matching}, pages 116--129, 1997.
\newblock \href {https://doi.org/10.1007/3-540-63220-4_55}
  {\path{doi:10.1007/3-540-63220-4_55}}.

\bibitem{Crochemore1997}
Maxime Crochemore and Renaud V{\'{e}}rin.
\newblock {On compact directed acyclic word graphs}.
\newblock In {\em Structures in Logic and Computer Science: A Selection of
  Essays in Honor of A. Ehrenfeucht}, pages 192--211. Springer, 1997.
\newblock \href {https://doi.org/10.1007/3-540-63246-8_12}
  {\path{doi:10.1007/3-540-63246-8_12}}.

\bibitem{Ehrenfeucht2011}
Andrzej Ehrenfeucht, Ross~M. McConnell, Nissa Osheim, and Sung-Whan Woo.
\newblock {Position heaps: A simple and dynamic text indexing data structure}.
\newblock {\em Journal of Discrete Algorithms}, 9(1):100--121, 2011.
\newblock \href {https://doi.org/10.1016/j.jda.2010.12.001}
  {\path{doi:10.1016/j.jda.2010.12.001}}.

\bibitem{Farach-ColtonFM00}
Martin Farach{-}Colton, Paolo Ferragina, and S.~Muthukrishnan.
\newblock On the sorting-complexity of suffix tree construction.
\newblock {\em Journal of the {ACM}}, 47(6):987--1011, 2000.
\newblock \href {https://doi.org/10.1145/355541.355547}
  {\path{doi:10.1145/355541.355547}}.

\bibitem{0001G15}
Johannes Fischer and Pawel Gawrychowski.
\newblock Alphabet-dependent string searching with wexponential search trees.
\newblock In {\em Proceedings of 26th Annual Symposium on Combinatorial Pattern
  Matching}, pages 160--171, 2015.
\newblock \href {https://doi.org/10.1007/978-3-319-19929-0\_14}
  {\path{doi:10.1007/978-3-319-19929-0\_14}}.

\bibitem{FujishigeTIBT16}
Yuta Fujishige, Yuki Tsujimaru, Shunsuke Inenaga, Hideo Bannai, and Masayuki
  Takeda.
\newblock Computing dawgs and minimal absent words in linear time for integer
  alphabets.
\newblock In {\em Proceedings of the 41st International Symposium on
  Mathematical Foundations of Computer Science}, pages 38:1--38:14, 2016.
\newblock \href {https://doi.org/10.4230/LIPIcs.MFCS.2016.38}
  {\path{doi:10.4230/LIPIcs.MFCS.2016.38}}.

\bibitem{Gabow1983}
Harold~N. Gabow and Robert~Endre Tarjan.
\newblock {A linear-time algorithm for a special case of disjoint set union}.
\newblock In {\em Proceedings of the fifteenth annual ACM symposium on Theory
  of computing}, pages 246--251, 1983.
\newblock \href {https://doi.org/10.1145/800061.808753}
  {\path{doi:10.1145/800061.808753}}.

\bibitem{Gabow1985}
Harold~N. Gabow and Robert~Endre Tarjan.
\newblock {A linear-time algorithm for a special case of disjoint set union}.
\newblock {\em Journal of Computer and System Sciences}, 30(2):209--221, 1985.
\newblock \href {https://doi.org/10.1016/0022-0000(85)90014-5}
  {\path{doi:10.1016/0022-0000(85)90014-5}}.

\bibitem{GasieniecKPS05}
Leszek Gasieniec, Roman~M. Kolpakov, Igor Potapov, and Paul Sant.
\newblock Real-time traversal in grammar-based compressed files.
\newblock In {\em Proceedings of Data Compression Conference 2005}, page 458,
  2005.
\newblock \href {https://doi.org/10.1109/DCC.2005.78}
  {\path{doi:10.1109/DCC.2005.78}}.

\bibitem{GawrychowskiKKL18}
Pawel Gawrychowski, Adam Karczmarz, Tomasz Kociumaka, Jakub Lacki, and Piotr
  Sankowski.
\newblock Optimal dynamic strings.
\newblock In {\em Proceedings of the 2018 Annual ACM-SIAM Symposium on Discrete
  Algorithms}, pages 1509--1528, 2018.
\newblock \href {https://doi.org/10.1137/1.9781611975031.99}
  {\path{doi:10.1137/1.9781611975031.99}}.

\bibitem{Hendrian2019}
Diptarama Hendrian, Takuya Takagi, and Shunsuke Inenaga.
\newblock {Online Algorithms for Constructing Linear-size Suffix Trie}.
\newblock In {\em Proceedings of the 30th Annual Symposium on Combinatorial
  Pattern Matching}, pages 30:1--30:19, 2019.
\newblock \href {https://doi.org/10.4230/LIPIcs.CPM.2019.30}
  {\path{doi:10.4230/LIPIcs.CPM.2019.30}}.

\bibitem{INIBT16}
Tomohiro I, Yuto Nakashima, Shunsuke Inenaga, Hideo Bannai, and Masayuki
  Takeda.
\newblock Faster lyndon factorization algorithms for {SLP} and {LZ78}
  compressed text.
\newblock {\em Theoretical Computer Science}, 656:215--224, 2016.
\newblock \href {https://doi.org/10.1016/j.tcs.2016.03.005}
  {\path{doi:10.1016/j.tcs.2016.03.005}}.

\bibitem{Imai1987}
Hiroshi Imai and Takao Asano.
\newblock {Dynamic orthogonal segment intersection search}.
\newblock {\em Journal of Algorithms}, 8(1):1--18, 1987.
\newblock \href {https://doi.org/10.1016/0196-6774(87)90024-1}
  {\path{doi:10.1016/0196-6774(87)90024-1}}.

\bibitem{InenagaHSTAMP05}
Shunsuke Inenaga, Hiromasa Hoshino, Ayumi Shinohara, Masayuki Takeda, Setsuo
  Arikawa, Giancarlo Mauri, and Giulio Pavesi.
\newblock On-line construction of compact directed acyclic word graphs.
\newblock {\em Discrete Applied Mathematics}, 146(2):156--179, 2005.
\newblock \href {https://doi.org/10.1016/j.dam.2004.04.012}
  {\path{doi:10.1016/j.dam.2004.04.012}}.

\bibitem{KarkkainenSB06}
Juha K{\"{a}}rkk{\"{a}}inen, Peter Sanders, and Stefan Burkhardt.
\newblock Linear work suffix array construction.
\newblock {\em Journal of the {ACM}}, 53(6):918--936, 2006.
\newblock \href {https://doi.org/10.1145/1217856.1217858}
  {\path{doi:10.1145/1217856.1217858}}.

\bibitem{Kucherov2013}
Gregory Kucherov.
\newblock {On-line construction of position heaps}.
\newblock {\em Journal of Discrete Algorithms}, 20:3--11, 2013.
\newblock \href {https://doi.org/10.1016/j.jda.2012.08.002}
  {\path{doi:10.1016/j.jda.2012.08.002}}.

\bibitem{LarssonFK14}
N.~Jesper Larsson, Kasper Fuglsang, and Kenneth Karlsson.
\newblock Efficient representation for online suffix tree construction.
\newblock In {\em {SEA} 2014}, volume 8504 of {\em Lecture Notes in Computer
  Science}, pages 400--411. Springer, 2014.
\newblock \href {https://doi.org/10.1007/978-3-319-07959-2\_34}
  {\path{doi:10.1007/978-3-319-07959-2\_34}}.

\bibitem{Manber1993}
Udi Manber and Gene Myers.
\newblock {Suffix Arrays: A New Method for On-Line String Searches}.
\newblock {\em SIAM Journal on Computing}, 22(5):935--948, 1993.
\newblock \href {https://doi.org/10.1137/0222058} {\path{doi:10.1137/0222058}}.

\bibitem{NarisawaHIBT17}
Kazuyuki Narisawa, Hideharu Hiratsuka, Shunsuke Inenaga, Hideo Bannai, and
  Masayuki Takeda.
\newblock Efficient computation of substring equivalence classes with suffix
  arrays.
\newblock {\em Algorithmica}, 79(2):291--318, 2017.
\newblock \href {https://doi.org/10.1007/s00453-016-0178-z}
  {\path{doi:10.1007/s00453-016-0178-z}}.

\bibitem{Takagi2017}
Takuya Takagi, Keisuke Goto, Yuta Fujishige, Shunsuke Inenaga, and Hiroki
  Arimura.
\newblock {Linear-Size CDAWG: New Repetition-Aware Indexing and Grammar
  Compression}.
\newblock In {\em Proceedings of the 24th International Symposium on String
  Processing and Information Retrieval}, pages 304--316, 2017.
\newblock \href {https://doi.org/10.1007/978-3-319-67428-5_26}
  {\path{doi:10.1007/978-3-319-67428-5_26}}.

\bibitem{Ukkonen1995}
Esko Ukkonen.
\newblock {On-line construction of suffix trees}.
\newblock {\em Algorithmica}, 14(3):249--260, 1995.
\newblock \href {https://doi.org/10.1007/BF01206331}
  {\path{doi:10.1007/BF01206331}}.

\bibitem{Weiner1973}
Peter Weiner.
\newblock {Linear pattern matching algorithms}.
\newblock In {\em Proceedings of the 14th Annual Symposium on Switching and
  Automata Theory}, pages 1--11. IEEE, 1973.
\newblock \href {https://doi.org/10.1109/SWAT.1973.13}
  {\path{doi:10.1109/SWAT.1973.13}}.

\end{thebibliography}
